%% file: 2017july1_local_figs.tex
\documentclass[]{spie}  
\usepackage[]{graphicx}
\usepackage{glossaries}
\usepackage{hyperref}
\usepackage{subcaption}
\title{Wavefront Sensing in Space: Flight Demonstration II of the PICTURE Sounding Rocket Payload} 

\author{Ewan S. Douglas\supit{a}, 
Christopher B. Mendillo\supit{b},
Timothy A. Cook\supit{b}, 
Kerri L. Cahoy\supit{a,c},
Supriya Chakrabarti\supit{b}
\skiplinehalf
\supit{a}Department of Aeronautics and Astronautics, Massachusetts Institute of Technology, Cambridge, MA, USA 02139; \\
\supit{b}Lowell Center for Space Science and Technology, UMASS Lowell, Lowell, MA, USA 01854;\\
\supit{c}Department of Earth, Atmospheric, and Planetary Science, Massachusetts Institute of Technology, Cambridge, MA, USA 02139; \\
}

\authorinfo{Further author information: (Send correspondence to E.S.D.)\\E.S.D.: E-mail: douglase@bu.edu}

\input{acronyms}

\begin{document} 
  \maketitle 

\begin{abstract}
A NASA sounding rocket for high-contrast imaging with a visible nulling coronagraph, the \gls{PICTURE} payload,  has made two suborbital attempts to observe the warm dust disk inferred around Epsilon Eridani.
The first flight in 2011 demonstrated a 5 milliarcsecond fine pointing system in space.
Despite several anomalies in flight, post-facto reduction of the first ten seconds of phase stepping interferometer data provides insight into the wavefront sensing precision and the system stability.
The reduced flight data  from the second launch on 25 November 2015, presented herein, demonstrate active sensing of wavefront phase in space with a precision of  2.1 $\pm$ 1.7 nanometers per pixel, a system stability of approximately  4.8 $\pm$ 4.2  nanometers per pixel, and activation of a 1020-actuator microelectromechanical system deformable mirror. 

\end{abstract}



\section{INTRODUCTION}
\label{sec:intro}  

Resolving reflected  light from planets in distant star systems analogous to our own  requires overcoming    exoplanet-host star flux ratios  between 10$^{-9}$ and  10$^{-11}$. 
Imaging at these extreme contrast ratios\cite{turnbull_spectrum_2006,cahoy_exoplanet_2010}  will be enabled by  coronagraphs on the next generation of space telescopes. 
Both internal and external coronagraphs block light from the host star and  transmit light from dim companions to the detector.
External coronagraphy employs a starshade flying far from the observing telescope to occult starlight before it enters the telescope aperture\cite{spitzer_beginnings_1962}, which requires precise fabrication of the starshade petals and continuous station-keeping between two spacecraft separated by hundreds of thousands of kilometers\cite{seager_exo-s:_2015}.
An internal coronagraph suppresses starlight between the primary mirror and the detector image plane of a space telescope.
Internal coronagraphy requires wavefront stability, and with stellar leakage typically proportional to the square of the wavefront error\cite{traub_direct_2010}, dynamic and static aberrations in wavefront phase must be controlled to nanometer levels to detect self-luminous giant planets and debris disks while detection of terrestrial planets requires sub-angstrom control\cite{guyon_theoretical_2006,trauger_laboratory_2007}.

In addition to these  and other  implementation challenges, circumstellar dust in extrasolar systems contributes  a bright background to exoplanet observations. 
Depending on the  optical density, which is unmeasured for even nearby systems, detection of exoplanets against this exozodiacal background may require longer observing times and increased spatial resolution\cite{roberge_exozodiacal_2012,turnbull_search_2012}.
Variation in the dust morphology may hint at the presence of planets \cite{stark_detectability_2008}.  
Conversely, morphology also increases confusion between exoplanets and dust  \cite{defrere_nulling_2010}. %
 Efforts are underway to measure exozodiacal dust populations in infrared emission,\cite{hinz_large_2003,eiroa_dust_2013} but  visible light observations of circumstellar exozodiacal light are necessary to directly constrain the background signal which must be overcome by future missions to image exoplanets in the wavelengths where reflected light is brightest.
 
\begin{figure}[htp]
   \caption{The PICTURE \gls{VNC} optical layout (top left), and a photograph of the  instrument with the optical path overlaid (bottom right).
   The input beam from the telescope is first divided by a 600 nm short-pass dichroic beamsplitter into the \gls{VNC} and \gls{FPS} angle tracker camera (AT) paths. 
   The \gls{VNC} arm is next split into two arms at beamsplitter 1. 
   The optical path difference between the sheared beams is matched by the \gls{NPZT} and \gls{DM}. 
   Interference occurs at beamsplitter 2, where the dark and bright fringes are split.
   In the flight configuration, the calibration interferometer was disabled and both the science (SCI) and \gls{WFS} cameras observed the dark fringe output via beamsplitter 3. }
   \centering  
   \includegraphics[width=0.8\textwidth]{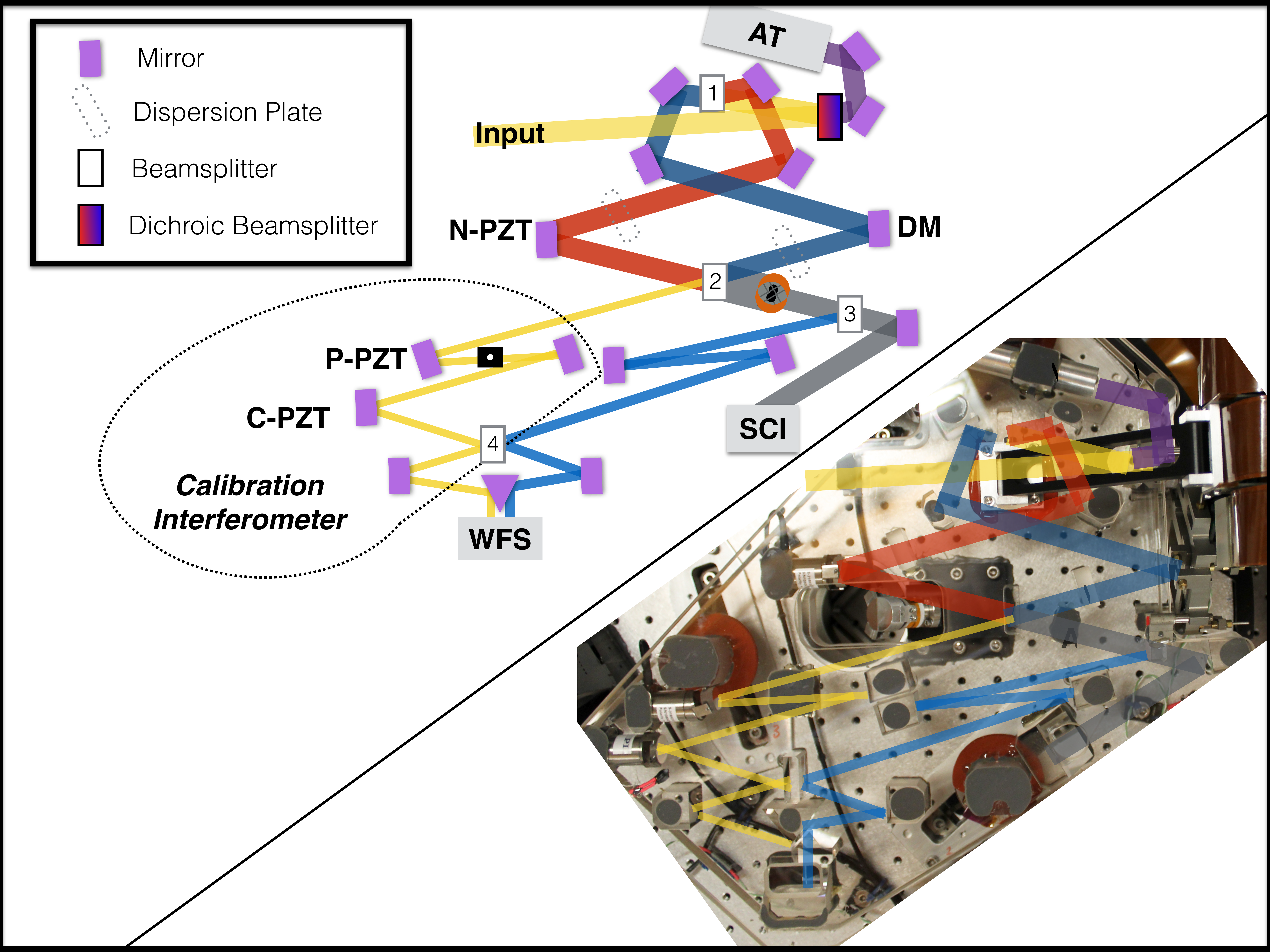}
      \label{fig:payloadschematic}
      \end{figure}

 The suborbital \gls{PICTURE}\cite{shao_nulling_2006,samuele_experimental_2007,rao_path_2008,mendillo_flight_2012,mendillo_picture:_2012,douglas_end--end_2015,chakrabarti_planet_2016} sounding rocket payload employs an internal \gls{VNC}, Fig. \ref{fig:payloadschematic}, behind a half-meter telescope with the aim of performing high-contrast observations of exozodiacal dust in space while demonstrating internal coronagraphy along with the requisite wavefront sensing and control.  
  
\subsection{Objectives}
 The \gls{PICTURE} sounding rocket observational objective was to measure the scattered light from exozodiacal dust around \gls{epseri}  at visible wavelengths.
\gls{epseri} has a large infrared excess at 20 $\mu$m which has been attributed to a dusty exozodiacal debris disk\cite{di_folco_near-infrared_2007}. 
With an expected integrated brightness of  approximately 2$\times10^{-4}L_\star$, this dust may be arranged either in a thin ring\cite{backman_epsilon_2009} or as a more diffuse debris disk populated by material streaming from the outer system via pseudo-Poynting-Robertson stellar wind drag\cite{reidemeister_cold_2011}.
 The \gls{PICTURE} missions planned to test for emission at separations from 2 AU to 20 AU, constraining  scattered light brightness and morphology and advancing our knowledge of the dust composition around sun-like stars. 
The predicted thin ring of emission provides a relatively bright target for demonstrating  high-contrast imaging, and the instrument technology development to achieve these objectives matures coronagraphy and wavefront sensing\cite{chakrabarti_planet_2016}. 
 
\subsection{Nulling Coronagraphy}\label{subsec:nulling}
In the  nulling coronagraph architecture\cite{bracewell_detecting_1978} (``nuller") two equal intensity beams of quasi-coherent starlight, collected by apertures separated by a baseline (interferometer ``arms'') with a relative phase shift of $\pi$ are combined to form a fringe pattern on the sky.
When recombination occurs at a beamsplitter, 
the output is divided into two paths:  the ``dark fringe" path where starlight  destructively interferes (``nulling''), and a second path where light constructively interferes, the ``bright fringe''.
Nulling coronagraphy requires coherent interference, which means the absolute path between each interferometer arm must be matched, otherwise the beams are temporally incoherent and interference fringes will not be observed. 
Within a few wavelengths of this absolute phase shift light  remains quasi-coherent, with the nuller  bandwidth in wavelength space defining the coherence length of this interference fringe packet 
The relative path differences between the two arms of the interferometer depend on the source angle on the sky with respect to the optical axis. 
Thus, when the fringe pattern is centered on a star, the light from exoplanets at close angles is only partially transmitted.
 Similarly, a small fraction of starlight is transmitted due to the finite size of the star\cite{serabyn_nulling_2000}. 
Nulling holds promise for high-contrast  exoplanet visible light imaging with complex segmented apertures\cite{hicks_high-contrast_2014} and mid-\gls{IR} detection with formation flying telescopes\cite{beichman_status_2006}.

The \gls{PICTURE} \gls{VNC} design\cite{rao_path_2008} is a uni-axial Mach-Zehnder \gls{LSI} design\cite{shao_visible_2004,shao_calibration_2005,lyon_visible_2006,levine_visible_2006} with dispersion plates which allow for broadband nulling\cite{morgan_nulling_2000}. 
Fig. \ref{fig:payloadschematic} shows that by splitting the input wavefront with a beamsplitter (labeled 1) and offsetting the two arms laterally, the \gls{LSI} design allows for interference between two sub-apertures formed from a single telescope pupil at the beamsplitter labeled 2.  
The first order nuller transmission pattern, $T(b,\theta,\lambda)$, is a function of the baseline, or shear length, between the two sub-apertures ($b$), the angle on the sky along the shear axis ($\theta$), and the wavelength ($\lambda$). 
The fringe pattern for a nulling coronagraph is analogous to the interference generated by  Young's double slit experiment\cite{angel_imaging_1997}.
Variation in amplitude or phase between the two beams results in partial transmission, or leakage $L$,  of on-axis starlight proportional to the square of the error\cite{serabyn_nulling_2000}. 
For small phase errors, $\Delta\phi<<1$,  the fractional starlight leakage into the image plane is approximately \cite{hicks_nulling_2012,douglas_advancing_2016}:
\begin{equation}
L_\phi\approx\frac{\Delta\phi^2}{4}.\label{eq:phase_leakage}
\end{equation}
Accurately measuring  and correcting these phase errors is essential to preventing leakage of starlight through the interferometer.

\subsection{Wavefront Control}

Space-based telescopes are unaffected by atmospheric turbulence, enabling diffraction limited imaging even at short wavelengths.
In space, time-varying \gls{WFE} primarily arises from the coupling of mechanical and thermomechanical perturbations of the spacecraft structure to the alignment and surfaces of optical components.
Directly imaging reflected light from exoplanets requires sensing and correcting sub-nanometer wavefront disturbances\cite{traub_direct_2010}.
Wavefront sensing of the  \gls{HST} telescope has relied on post-facto phase-retrieval from images \cite{krist_phase-retrieval_1995}.
The Gaia mission employs a passive Shack-Hartmann wavefront sensor\cite{vosteen_wavefront_2009} and a wide variety of new wavefront sensing techniques
are being developed for future space telescopes \cite{feinberg_trl-6_2007,greenbaum_-focus_2016,shi_low_2016}.
Active wavefront phase sensing is implemented in the  \gls{PICTURE} \gls{VNC} using phase stepping interferometry\cite{wyant_use_1975}, where the intensity is measured at multiple points across the fringe pattern by changing the relative path length between two arms.

In the \gls{PICTURE} \gls{VNC}, an additional post-coronagraph calibration interferometer allows sensing of residual errors while nulling by interfering the dark and bright outputs of the \gls{VNC}  (beamsplitter 3 in Fig. \ref{fig:payloadschematic})\cite{rao_path_2008}.
 This calibration interferometer was not employed in the refurbished payload on the ground or in-flight.

\subsection{Deformable Mirrors}
Deformable mirrors enable precision wavefront control by dynamically changing the optical path length across a range of  spatial frequencies. 
The \gls{PICTURE} \gls{VNC} employs a \gls{DM} to minimize the phase \gls{WFE} between the two arms of the interferometer, enabling the nulling of starlight even in the presence of wavefront error between the sheared pupil sub-apertures. 
Numerous deformable mirror technologies exist or have been proposed, including piezoelectric\cite{ealey_xinetics_1994}, thermoelectric coolers\cite{huang_experimental_2015}, ferrofluid \cite{lemmer_mathematical_2016}, and \gls{MEMS}\cite{bifano_microelectromechanical_1999}.
Compact size, high actuator count, low power consumption, and extensive use in ground-based adaptive optics\cite{morzinski_mems_2012}  make \gls{MEMS} deformable mirrors particularly desirable for space-based applications.

\gls{MEMS} deformable mirrors typically rely on voltages up to 250 V to electrostatically displace a membrane on sub-angstrom to micron scales\cite{bifano_microelectromechanical_1999}. 
There are few references to \gls{MEMS} optical device operation in space.
Yoo et al.\cite{yoo_mems_2009} found a non-deformable (on-off only) \gls{MEMS} micromirror device maintained  functionality after undergoing launch and was successfully operated on the \gls{ISS}.
A magnetically actuated \gls{MEMS} microshutter array has been flown on the Far-ultraviolet Off Rowland-circle Telescope for Imaging and Spectroscopy (FORTIS) sounding rocket\cite{fleming_calibration_2013}.
Finite element modeling has been used to predict the survival of \gls{MEMS} \gls{DM}s subjected to launch\cite{aguayo_fem_2014}, but survival in a harsh launch environment has not been previously demonstrated in the literature. 
                                    
The \gls{PICTURE} \gls{DM} is a Boston Micromachines 32 $\times$ 32 square Kilo-DM with a 1.5 $\mu$m stroke, a continuous gold-coated membrane, a 340 $\mu$m actuator pitch, and custom drive electronics\cite{rao_path_2008}. 
Actuator drive voltages are limited to $\leq$ 150 V to prevent actuator snap down\cite{morzinski_mems_2012}.
 Some amount of the stroke is typically lost to correct residual stresses in the surface of a \gls{MEMS} \gls{DM}\cite{bifano_micromachined_2000}.
To avoid chromatic optical path mismatch between the two \gls{VNC} arms, the \gls{DM} lacks the protective window commonly included to prevent  mirror surface  contamination.

\subsection{Previous PICTURE Flight}

Flight I of the \gls{PICTURE} payload launched\cite{mendillo_flight_2012,mendillo_picture:_2012,hicks_nulling_2012,mendillo_scattering_2013} aboard \gls{nasa} sounding rocket 36.225 UG (a Black Brant IX University Galactic Astronomy mission) on 8 October 2011 from \gls{wsmr}. 
The  flight  suffered a telemetry failure approximately seventy seconds after launch during calibration observations of Rigel.
Data recorded onboard showed the \gls{FPS} demonstrated first order \gls{AO},  sensing and controlling tip and tilt errors to approximately $5\times10^{-3}$ arcseconds \gls{RMS} at an update rate of 200 Hz to remove  residual attitude control system jitter \cite{mendillo_flight_2012}.
Unfortunately, limited \gls{WFS} data was transmitted before telemetry failure and no interference fringes were observed, preventing wavefront sensing or nulling.
A second mission,  renamed PICTURE-B (Planet Imaging Coronagraphic Technology Using a Reconfigurable Experimental Base) refurbished and relaunched the same payload with minor modifications in the fall of 2015 with the same science goals as the original flight is discussed in this paper.

\subsection{Payload Reflight as PICTURE-B}
The flight data presented herein was collected during flight of the refurbished payload, \gls{PICTURE}-B (NASA 36.293 UG), launched from \gls{wsmr} at 9:17 p.m. MST November 24th (25 November  2015 0417 UT).
The payload design, concept of operations, and the Flight I telemetry failure were described previously by Mendillo et al.\cite{mendillo_picture:_2012}, and refurbished telescope and nulling coronagraph integration and performance were summarized by Chakrabarti et al\cite{chakrabarti_planet_2016}. 
 Fortunately, the Flight II (36.293) telemetry system performed as designed, and data were redundantly stored onboard, providing far more insight into instrument performance in flight.
In this work we focus on the operation of the \gls{WFS} and \gls{DM} in space during the second flight.
  Section \ref{sec:refurb},  briefly summarizes the refurbishment of the payload.
  Section \ref{sec:observations},  describes the flight observation sequence and the anomalous \gls{WFS} measurements recorded in flight.
     Section \ref{sec:methods},  details the post-processing methods used to interpret the flight data,
    Section \ref{sec:results},  presents the estimated  precision of these measurements. 
  Finally,  Section  \ref{sec:conclusion}, remarks on conclusions and future directions.

\section{Refurbishment}\label{sec:refurb}
\subsection{Deformable Mirror}
The cabling to the DM was damaged during the assembly of the payload for Flight I.
Thus, a new polyimide flex cable assembly was manufactured and installed along with a replacement Boston Micromachines Corporation  Kilo-DM (S.N. 11W310\#002).
This new \gls{DM} has two inactive actuators (a 99.8\% yield). 
Fortunately,  both inactive actuators were positioned behind the Lyot mask, which blocks un-interfered light from behind the sheared secondary obscurations, allowing active phase control across the entire output pupil.
To best match reflectivity between the interferometer arms, a new \gls{NPZT} mirror,  coated in the same chamber as the replacement \gls{DM}, was also installed. 
Located in the \gls{VNC} arm opposite the \gls{DM}, the \gls{NPZT} mirror is mounted on a Physik Instrumente  S-316 piezoelectric stage which corrects \gls{TTP} errors between the two interferometer arms with piston range of approximately 8.5 $\mu$m.
The two optics were aligned in the laboratory such that the \gls{NPZT} mirror flattened the wavefront error in the \gls{VNC} at room temperature in the middle of the piezoelectric stage's range.

\subsection{Primary Mirror}
The original \gls{PICTURE} primary mirror\cite{antonille_figure_2008}, flown on  Flight I, did not survive reentry and landing.
A new light-weighted silicon carbide primary mirror with a silicon cladding was designed by AOA Xinetics/Northrop Grumman to survive the rigors of launch and provide a sufficiently low surface error  to demonstrate the \gls{VNC} in space and measure the predicted \gls{epseri} inner warm dust ring.
Laboratory alignment and testing at 1g predicted $\lambda/2$ \gls{PV} wavefront error in free-fall\cite{chakrabarti_planet_2016}.

\subsection{Preflight Testing}
The \gls{VNC} was tested post-refurbishment, without the telescope, using a simulated point source and a retro-reflecting mirror.
The contrast was found to be comparable to previous tests with central star leakage of approximately $10^{-3}$\cite{rao_path_2008,chakrabarti_planet_2016}.
The \gls{VNC} residual wavefront error was controlled to 5.7 nm with a 1$\sigma$ error of 2.6 nm\cite{douglas_advancing_2016}.
The fully assembled payload was tested end-to-end on a vibration suppressing optical table and nulling was observed \cite{chakrabarti_planet_2016}.
The leakage ratio in end-to-end testing with the complete telescope was  limited  to only a factor of a few by environmental disturbances such as atmospheric turbulence and optical bench vibration.

The integrated payload was shake-tested at \gls{wsmr} to Vehicle Level Two random thrust vibration\cite[Table 6.3.4-1]{noauthor_sounding_2015}. 
The sounding rocket payload assembly provides some damping and the integrated acceleration of the nuller assembly during the random vibration was 10.4 g \gls{RMS} versus the  12.7 g \gls{RMS} input. 

\section{Observations}\label{sec:observations}

\subsection{Target Acquisition}
Both Flight I and II used  Rigel  ($\beta$ Orionis, $m_v$=0.13) as the initial calibration star. 
Unfortunately, neither flight successfully advanced from Rigel to \gls{epseri}.
During ascent, the \gls{FPS} computer was powered on at t+47. 
The \gls{WFCS} and  telemetry processing computer was powered on at t+50 seconds.
The \gls{FPS} camera controller was powered on at t+74 seconds.
After initial   acquisition of Rigel by the \gls{wff} \gls{acs}, the payload Xybion\textregistered \ camera, with an approximately 10 arcminute circular \gls{FOV}, was used to measure the pointing offset between the angle tracker \gls{ccd}  (Fig. \ref{fig:payloadschematic})  and the \gls{acs} system. 
During both missions, a manual uplink successfully provided the pointing correction to the \gls{acs}, placing Rigel on the angle tracker camera near the nominal t+105 second observing start time.
 Additional  uplinks commanded centering of the star on the angle tracker to the \gls{acs} accuracy of approximately 1 arcsecond.
Once Rigel was centered, the \gls{FPS}  control loop locked, providing pointing precision error of approximately five milliarcseconds, consistent with Flight I\cite{mendillo_flight_2012}. 
Flight I (36.225) did not advance past initial pointing due to the failure of a relay in the onboard telemetry system\cite{yuhas_sounding_2012}. 
After acquisition and \gls{FPS} lock, three attitude control system maneuvers were planned:
1) Nuller alignment and 10 seconds of speckle observations on Rigel 
2) Slew to \gls{epseri} and observe the circumstellar environment, 
3)  Roll payload 90 degrees during the \gls{epseri} observation to characterize speckles.
Flight II did not complete nuller alignment, but far more on-star observations were recorded.

\subsection{Data Products}\label{sec:data_products}
Two cameras observed the \gls{VNC} output after the Lyot stop, which transmitted only regions where the two interferometer arms overlap, the science camera (labeled SCI in Fig. \ref{fig:payloadschematic}) to image the sky and the \gls{WFS} camera to image the interference fringes in the pupil plane.
Cut-off filters limited the   observed bandwidth of both cameras to between 600 nm and 750 nm.
The  \gls{WFS} and science cameras were \gls{ccd} detectors developed for the Astro-E2 X-ray Imaging Spectrometer\cite{bautz_progress_2004}. 
These 1024$\times$1024 pixel MIT Lincoln Laboratory model CCID41 detectors were cryogenically cooled to -70$^\circ$C for a dark noise of approximately 1 e$^-$/s/pixel. 
In order to allow short exposure times, only small subregions were read-out from each camera. 
The \gls{WFS} readout area was 76 pixel $\times$ 76 pixel.
The integration time of each \gls{WFS} frame time was 0.23 seconds.
The laboratory measured read noise rate was 2.3 e$^-$ per pixel per exposure. 
 
\subsection{Nuller Alignment}
During Flight II the \gls{WFCS} advanced to the \gls{VNC} \textit{coarse mode} or ``phase-up" stage: locating the  fringe packet, applying a predetermined set of   voltages to  the \gls{DM},   to remove the stress-induced surface concavity from fabrication, and begin to flatten the wavefront error by eliminating optical path differences in \gls{TTP} between the interferometer arms by moving the \gls{NPZT} mirror.
In the planned flight sequence, coarse mode was followed by fine correction of higher spatial frequencies with the \gls{DM} and finally a transition to \textit{nulling mode} with the \gls{NPZT} shifted to the dark fringe for  high-contrast  science observation over the remainder of the flight.

Flight II did not reach the fine-mode correction or nulling modes because the wavefront could not be flattened, likely due to shift in the \gls{DM} mount, as discussed in Sec. \ref{sec:discussion}. 
The closed loop correction of wavefront mismatch between the arms with the \gls{DM} was not initiated because  the fringe packet was outside the range of the \gls{NPZT} to correct \gls{TTP} errors, keeping the fringe visibility  below the predetermined threshold to advance modes (Section \ref{eq:visibility}).  
The magnitude of the misalignment is estimated in Section \ref{sec:discussion}.

Since the system never entered nulling mode, the recorded science camera images are saturated and provide limited information about the interferometer state. 
The coarse flattening mode which was reached required accurate measurements of the wavefront phase error.
Thus, the mission did return  measurements of phase error and \gls{WFS} stability;  and the remainder of this analysis will focus solely on the \gls{WFS} camera data.

\subsection{Calculating the phase}\label{subsec:phasecalc}

 Simplifying the interference equation\cite{born_principles_1980} by assuming two beams of equal intensity ($I$) gives a relation between the phase difference, $\Delta\phi$, and the fringe intensity, $I(\Delta\phi$) between the beams:
  \begin{equation}\label{eq:intensity_phase}
  I(\Delta\phi)=2I+2I\cos(\Delta\phi)\mu.
  \end{equation}
  Here $\mu$ is the coherence between the two beams. 
    $\mu$ is near unity for measurements at the center of the interference fringe packet.
The total phase difference can be written as $\Delta\phi=\delta+\Delta\phi'$ where $\Delta\phi'$ is the phase error we seek to measure and $\delta$ is a known relative phase step between separate measurements.
This allows expansion of the cosine term: $\cos(\Delta\phi)=\cos\delta\cos\Delta\phi'-\sin\delta\sin\Delta\phi'$. 
Defining three new variables allows us to simplify the relation, $a_0=2I$, $a_1=a_0\cos\Delta\phi'$, and $a_2=-a_0\sin\Delta\phi'$, such that: $I(\Delta\phi)=a_0+a_1\cos\delta+a_2\sin\delta$.

 The \gls{PICTURE} payload was designed to recover phase by recording \gls{WFS} intensity measurements as a sequence of four measurements separated by $\pi/2$. 
 For convenience, we rename each of these intensities: \textbf{A}$=I(\delta=0)$,  \textbf{B}$=I(\delta=\pi/2)$,  \textbf{C}$=I(\delta=\pi)$,  \textbf{D}$=I(\delta=3\pi/2)$. 
Solving the system of equations composed of the four intensity measurements and the known phase step values  permits calculation of the phase error of each pixel in a set of ABCD measurements\cite{wyant_phase-shifting_2011,wyant_use_1975}:

\begin{equation}\label{eq:phi_ABCD}
\Delta\phi' = \arctan(\frac{A-C}{B-D}).
\end{equation}

Interference fringes in intensity due to $\Delta\phi'$ are visible when the path lengths are matched to within the coherence length of the fringe packet. 
For the 150 nm bandpass the coherence length at 675 nm is approximately 3 $\mu$m\cite[p. 320]{born_principles_1980}. 
\begin{figure}[htp]
   \caption{Visibility versus frame number for flight observations of Rigel.
    Each point represents the median visibility across all illuminated pixels of a  \gls{WFS} measurement. 
    The  horizontal line represents the lock state of the \gls{FPS}, with breaks in the line indicating times when the telescope was repointed, which caused the visibility to vary. 
The shaded  gray region represents the period used to calculate \gls{WFS} performance. 
In the later \gls{FPS} locked periods, the telescope focus had deteriorated, which decreased the \gls{FPS} stability and consequently increased the variance in the visibility. 
}
   \centering  
   \includegraphics[width=0.6\textwidth]{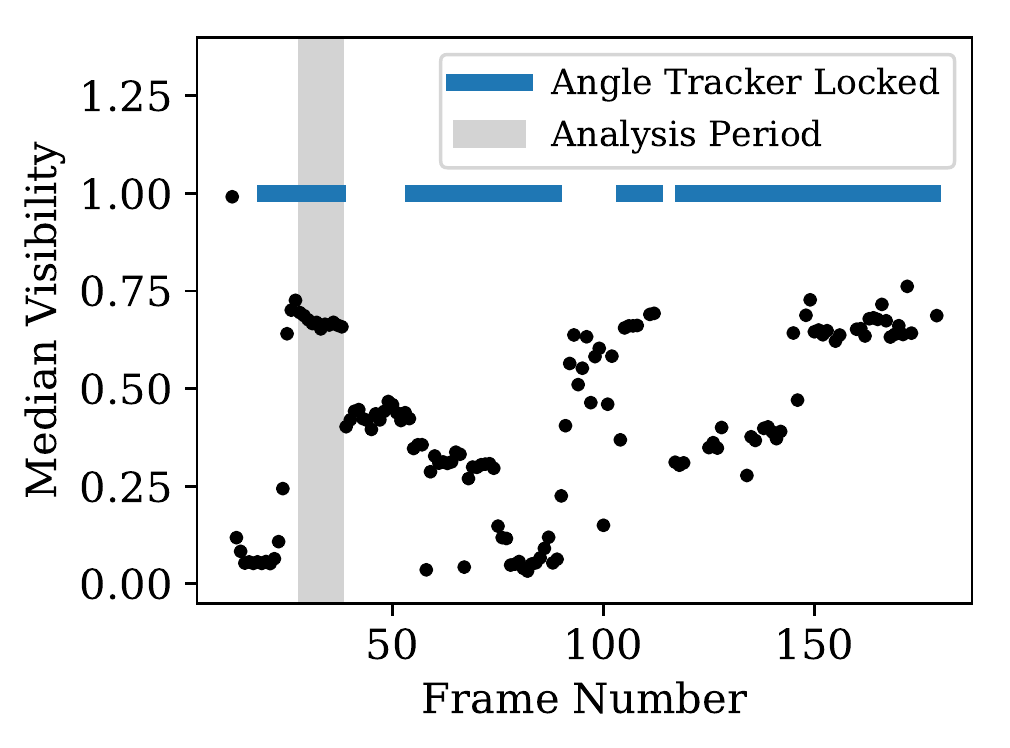}
      \label{fig:visibility_vs_time}
\end{figure}

Fringe visibility\cite{born_principles_1980}, $V$, a measure of the degree of coherence in an interference pattern, is expressed as:
 \begin{equation}
V\equiv \frac{I_{max}-I_{min}}{I_{max}+I_{min}}.\label{eq:visibility}
\end{equation}
When $V=1$ there is complete interference and cancellation of on-axis light at the center of the fringe packet while $V=0$ indicates completely incoherent light. 
Black dots in Fig. \ref{fig:visibility_vs_time} illustrate \gls{VNC} performance as measured by the median \gls{WFS} visibility. 
The cadence between each set of four images is approximately one second.
A visibility $>0.9$ across the pupil was expected; however, the median flight visibility never exceeded 0.8 due to incomplete steps of the \gls{NPZT} mirror, as will be described in  Sec. \ref{subsec:example_WFS}.
The highest visibility  was recorded approximately thirty seconds into the observation, when the first string of measurements were recorded.
 Rigel continued to be observed for the remainder of the flight. 
 Before the telescope shutter closed for reentry, several unsuccessful attempts were made to repoint the payload via human-in-the-loop command uplinks from the ground to bring the \gls{NPZT} into piston range. 
 This caused low visibility during each pointing maneuver and  re-acquisition of Rigel by the \gls{FPS}. 
  Closed loop \gls{FPS} control was lost during repointing, as seen in the ``lock'' status of the \gls{FPS} angle tracker.
  This status is plotted as a horizontal line in Fig. \ref{fig:visibility_vs_time}.
  The futility of these attempts is seen in the increased variability of the visibility as a function of time, as telescope focus deteriorated and the coherent fringe packet drifted further out of the \gls{NPZT} range.
The focus degraded due to thermal gradients across the telescope optical bench (see Douglas\cite{douglas_advancing_2016} for details on the thermal environment and the \gls{PSF} measured by the \gls{FPS} camera).

\begin{figure}[htp!]
\begin{centering}
	\begin{subfigure}[t]{\textwidth}
       			\includegraphics[width=\textwidth]{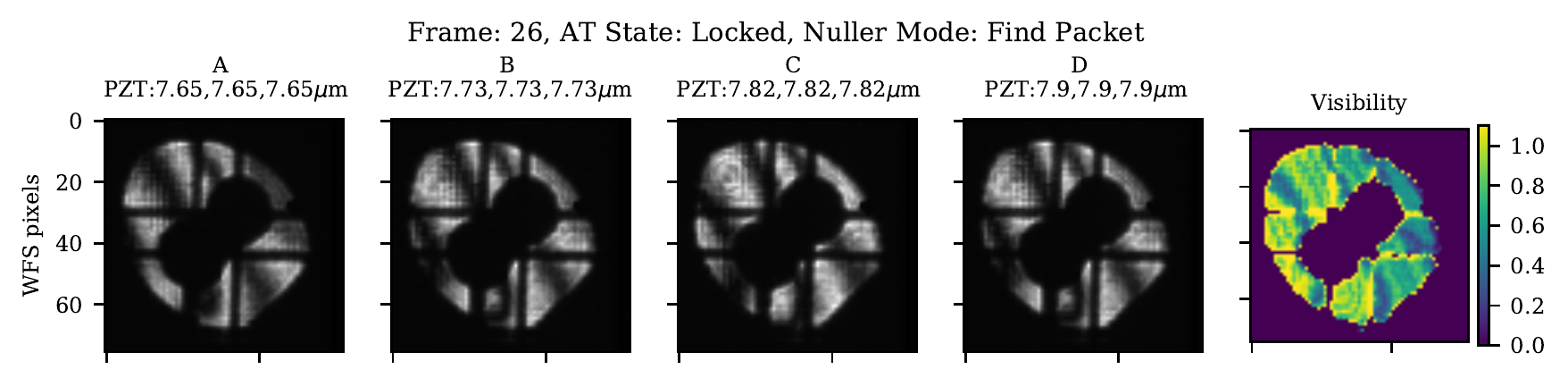}	
				\caption{Example measurement with complete phase steps and the \gls{NPZT} mirror in the initial flat position, leaving a large tilt relative to the input beam. 
				Since each step is complete, the fall-off in visibility across the pupil is due to the finite extent of the fringe packet. }\label{fig:ABCD26}
       \end{subfigure}
	\begin{subfigure}[t]{\textwidth}
       			\includegraphics[width=\textwidth]{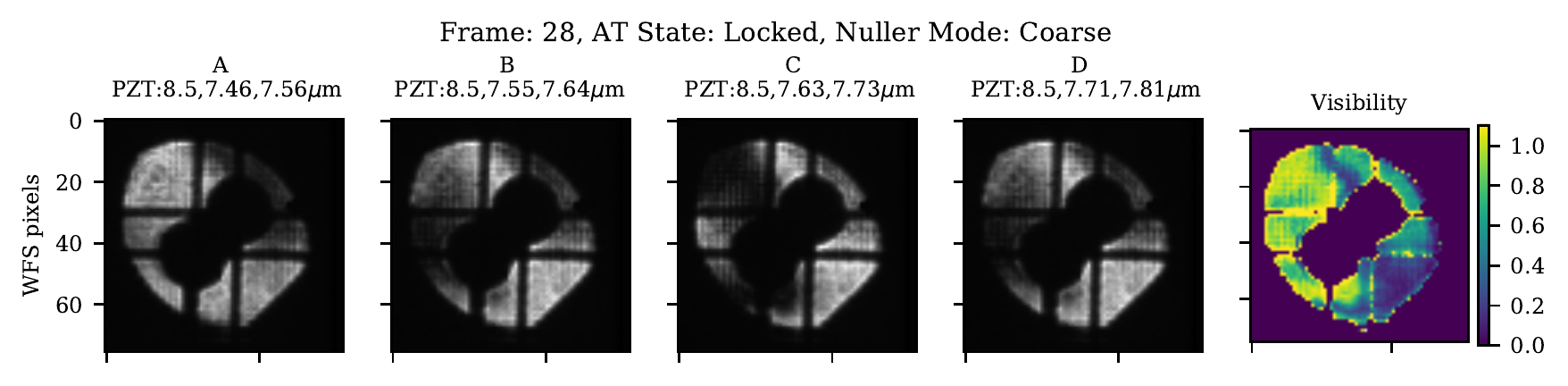}
		\caption{The first measurement recorded in coarse alignment mode shows the \gls{WFCS} was unable to flatten the fringes. 
		The lower visibility pixels on the right side of the pupil vary less between intensity images because a \gls{NPZT} actuator was railed.}\label{fig:ABCD28}
       \end{subfigure}
       	\begin{subfigure}[t]{\textwidth}
       			\includegraphics[width=\textwidth]{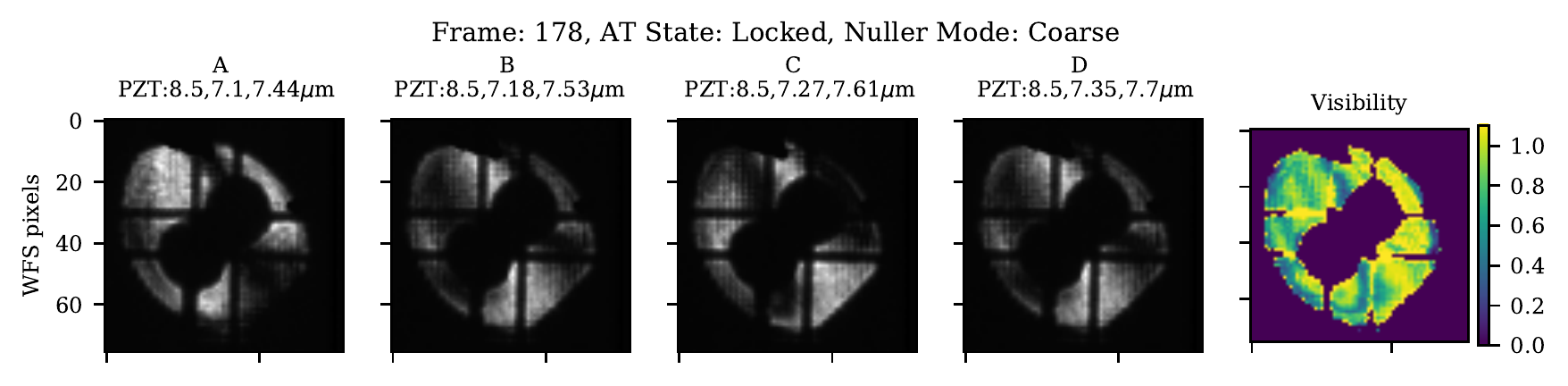}	\caption{The final \gls{WFS} measurement with the pointing system locked and the \gls{NPZT} mirror at an angle close to Fig. \ref{fig:ABCD28}.
	}\label{fig:ABCD178}
       \end{subfigure}
          \caption{ Examples of raw  wavefront sensor intensity measurements during Flight II of the star Rigel.
    The first four grayscale columns correspond to the four \gls{NPZT} positions A, B, C, and D, each shifted by a quarter-wave. 
    The position of the each of the three \gls{NPZT} actuators is shown in microns in the title of each image, these motions correspond to half of the relative wavefront shift, $\delta$.
    The first actuator listed is unchanging and railed at 8.5 $\mu$m for each image in the top and bottom rows.
    The far right column shows the visibility of each row of measurements.
    } 
   \label{fig:ABCDflight}
   \end{centering}

\end{figure}

\subsection{Example Wavefront Sensing Data}\label{subsec:example_WFS}
The three sets of quarter-wave step \gls{WFS} images shown in Fig. \ref{fig:ABCDflight} typify the range of raw measurements recorded over the course of the flight.
Since the wavefront was not flattened, fringes are visible across the pupil plane images recorded by the \gls{WFS} camera and  provide additional insight into the alignment of the \gls{VNC}. 
Each row  of Fig. \ref{fig:ABCDflight} is an example set of ABCD measurements from different times in the flight.
The left four columns illustrate the background subtracted intensity with \gls{WFS} frames taken at each of the the ABCD \gls{NPZT} positions.
(Background noise levels were calculated from the median of \gls{WFS} exposures recorded in-flight  before and after the Rigel observation.)  
The visibility measured from each sequence is shown in the rightmost column.
Fig. \ref{fig:ABCD26} shows a low visibility measurement where the \gls{NPZT} was functioning normally but was far from the center of the coherent fringe packet because the fringe flattening tip-tilt correction has not yet been applied.
Fig.  \ref{fig:ABCD28} shows the first set of high visibility measurements recorded at the beginning of the flight while the flight software was attempting to flatten the initial set of fringes across the pupil, the analysis period shaded in Fig. \ref{fig:visibility_vs_time}. 
The visibility on the left side of the pupil is high, but decreases towards the right side because the \gls{NPZT} actuator in that corner is out of range and ``railed'' at the maximum displacement.
Fig.  \ref{fig:ABCD28}  is representative of the series of  coarse alignment images where one of the three \gls{NPZT} actuators is railed at the beginning of the flight which we will use for the bulk of our analysis. 
Fig.  \ref{fig:ABCD178} shows an ABCD image from the final sequence at the end of the flight, after the telescope had been repointed to change the path length between the arms. 
Defocus appears as a wavefront tilt between the \gls{DM} and \gls{NPZT} arms of the \gls{VNC}, keeping the difference between the interferometer arms beyond the correction range of the \gls{NPZT}.

\section{Methods}\label{sec:methods}
\subsection{Wavefront Sensing}\label{sec:methods_wfs}
The \gls{PICTURE} design leverages the interferometric nature of a nulling coronagraph to directly measure wavefront error by imaging the pupil at the science output of the \gls{VNC}.
Before nulling, the wavefront into the system must first  be measured and corrected to flatten the interference fringe.
This section describes issues encountered in flight, and post-facto analysis to retrieve the phase.

The flight \gls{VNC} control software expected $\pi/2$ phase steps and
used Eq. \ref{eq:phi_ABCD} to calculate the phase error to be corrected by the \gls{NPZT} and \gls{DM}\cite{mendillo_picture:_2012}.
In order to better estimate the uncertainties in the returned data given the anomalous \gls{NPZT} steps,  raw intensity maps are re-reduced to find the best-fit phase in each pixel.

\subsubsection{Least Squares Fitting of the Phase}
Unfortunately, during Flight II one of the three piezo actuators translating the \gls{NPZT} mirror was railed high for many of the ABCD measurements, while the other two actuators moved the mirror in $\pi/2$ steps, causing an varying phase shift ($\delta$) across the pupil image. 
This is particularly true of the high visibility measurements (e.g. Fig. \ref{fig:ABCD28} and Fig. \ref{fig:ABCD178}) where the path length between the arms was best matched, meaning the railed images are also the measurements with the most coherent interference.

\begin{figure}[htp]
   \centering  
   \includegraphics[width=.6\textwidth]{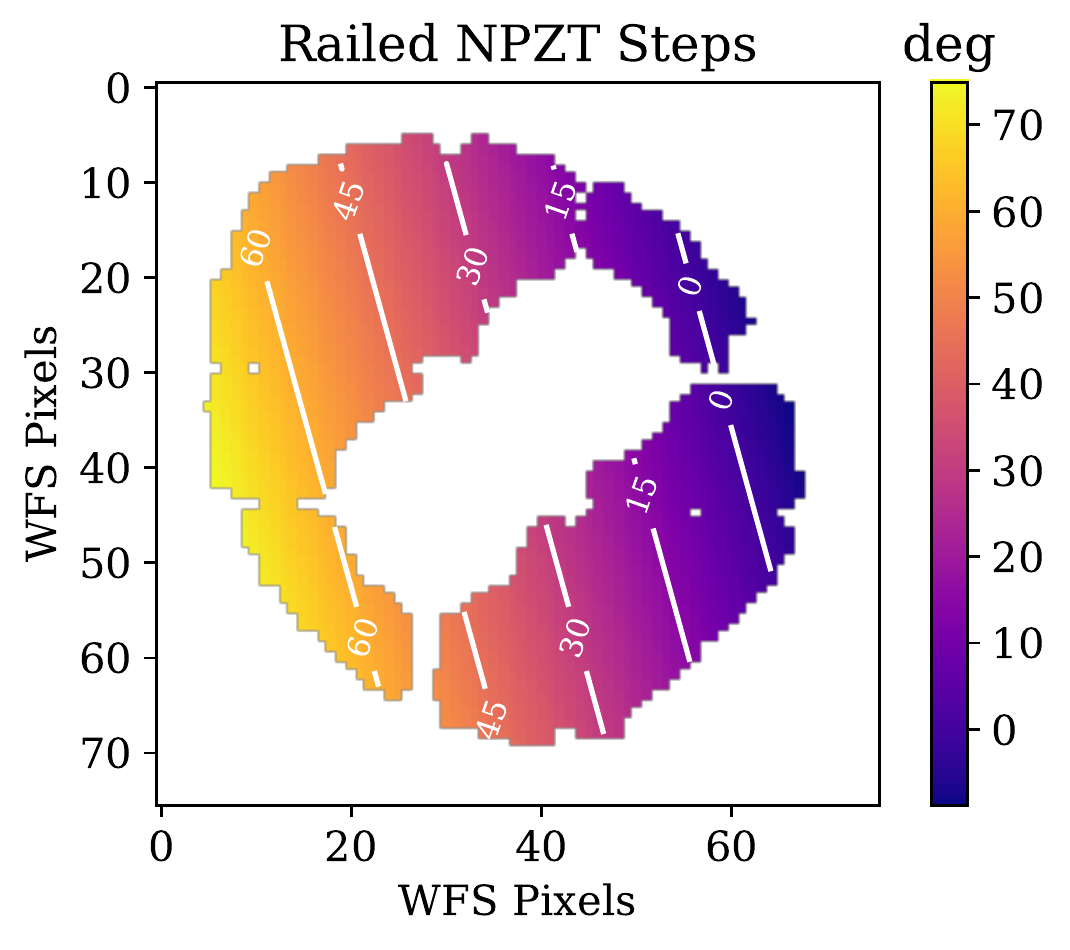}
      \caption{Example of an \gls{NPZT} phase step ($\delta$) with one actuator driven to its fullest extent  preventing a full step. 
   The resulting phase shift between steps can be seen to pivot about the railed actuator, with the largest shift on the left side of the pupil map. }
   \label{fig:RailedStepMap}
\end{figure}

To compensate for this uneven shifting of the \gls{NPZT} mirror,  an alternative approach to measuring phase was applied.
For varying values of $\delta$, the phase error ($\Delta\phi'$) can be recovered by least squares fitting of the intensity ($I(\Delta\phi)$) versus phase step-size ($\delta$). 
Allowing for variation in coherence, we again expand Eq. \ref{eq:intensity_phase} and fit  a model of three unknowns:
\begin{equation}
I(\Delta\phi)=a_0+(a_0\cos\Delta\phi'\cos\delta-a_0\sin\Delta\phi'\sin\delta)\mu.
\end{equation}
$\delta$ values for each \gls{WFS} pixel were calculated from commanded \gls{NPZT} positions using a laboratory calibrated transformation matrix of \gls{NPZT} data values to the \gls{TTP} values in units of distance.
 An example map of the $\delta$ step-size during flight with one railed actuator is shown across the \gls{WFS} pupil plane in Fig. \ref{fig:RailedStepMap}. 

\begin{figure}[t]
   \caption{Wrapped (left) and unwrapped (middle) pupil plane phase measurements from least squares fitting of four wavefront sensor measurements and corrected \gls{NPZT} positions.
  The error (right panel) shows $1\sigma$ fitting error including photon noise. 
  The phase error measurement rapidly deteriorates once the phase step across the pupil drops below $\pi/4$ due to the railed actuator.}   \label{fig:lsq_phi_maps}
   	\includegraphics[width=\textwidth]{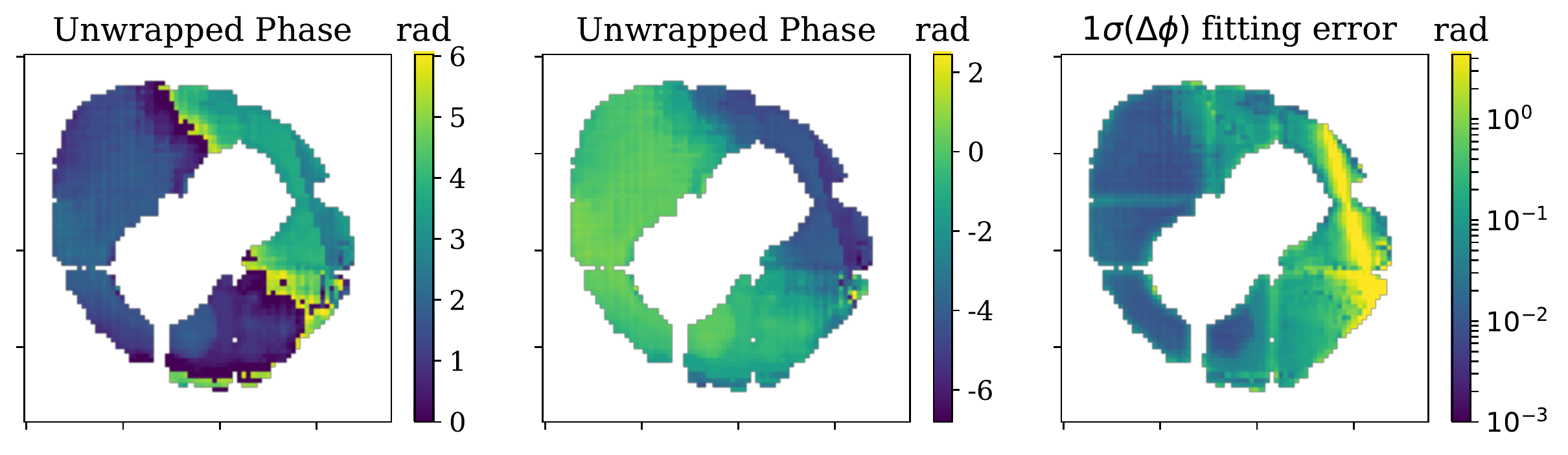}

\end{figure}
To constrain the problem, bounds were set requiring a coherence between $1\times10^{-9}$ and unity and a phase shift between $0$ and $2\pi$. 
This least squares bound-constrained minimization was solved using the subspace trust region interior reflective algorithm\cite{branch_subspace_1999} implemented in SciPy 0.19\cite{jones_scipy:_2001}. 
Least squares fitting of each pixel was repeated on the four images of a railed measurement with varying values of $\delta$. 
An example of the resulting phase map is shown in Fig. \ref{fig:RailedStepMap}. 

The phase measurements wrap about $2\pi$ radians. 
These measurements were unwrapped in order of pixel reliability in a noncontiguous fashion via the Herr\'{a}ez\cite{herraez_fast_2002} method.
 The middle panel of Fig. \ref{fig:lsq_phi_maps} shows an unwrapped phase measurement.

\begin{figure}[htp]
   \centering  
\begin{subfigure}[t]{0.45\linewidth}
   \includegraphics[width=.9\textwidth]{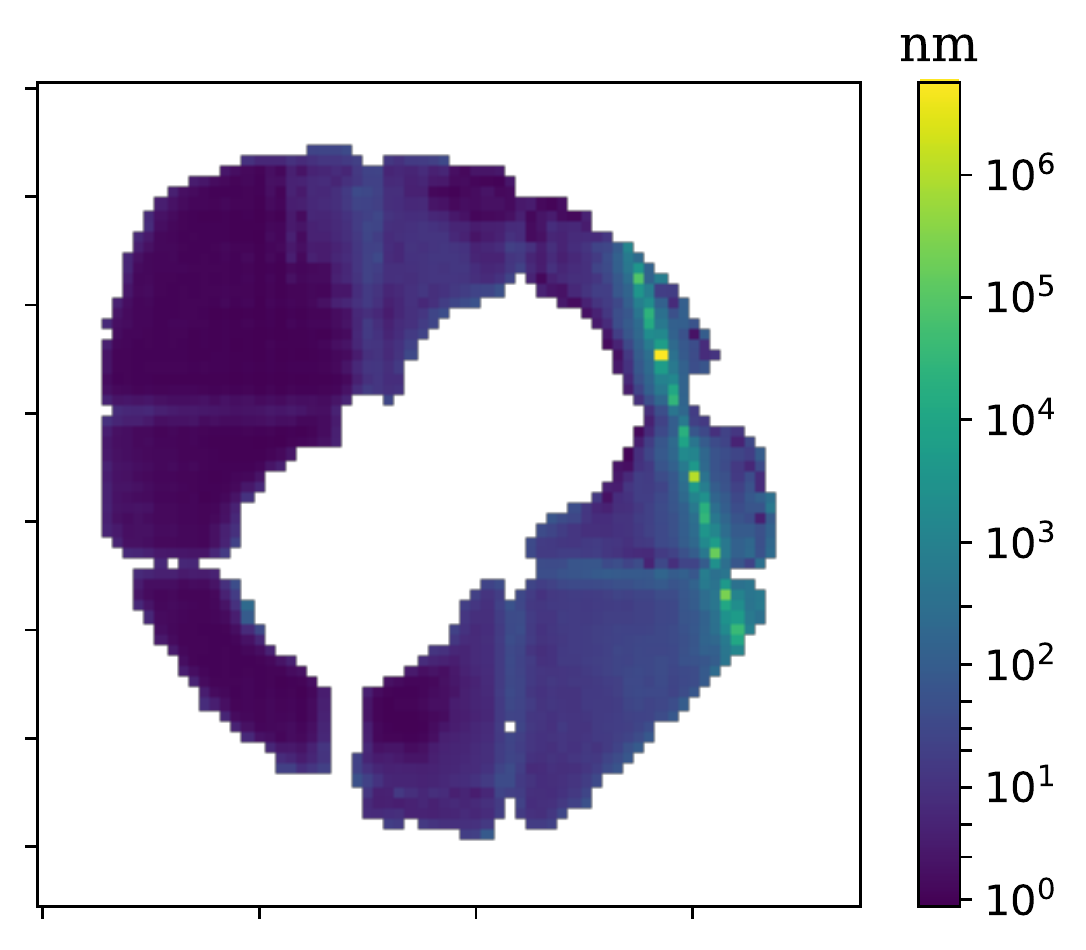}
     \caption{The mean fitting error across the wavefront sensor. 
     The uncertainty peaks along the axis where the railed \gls{NPZT} actuator held the mirror pinned.}  \label{fig:lsq_avg_uncertainty}
     \end{subfigure}
                             \hfill          
\begin{subfigure}[t]{0.45\linewidth}
   \includegraphics[width=.9\textwidth]{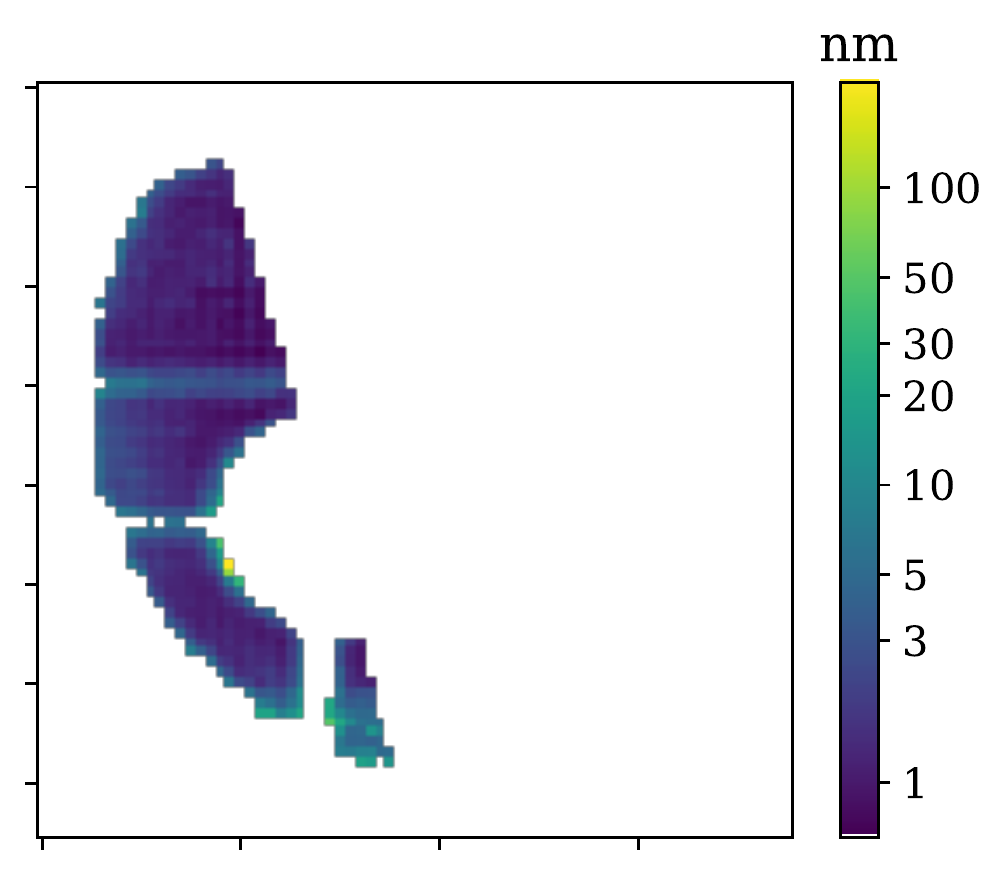}
     \caption{The same as \ref{fig:lsq_avg_uncertainty}, except only showing those pixels where the \gls{NPZT} stepped more than $\pi/4$.  }\label{fig:lsq_avg_uncertainty_good}
     \end{subfigure}
  \begin{subfigure}[t]{0.45\linewidth}
     \includegraphics[width=.9\textwidth]{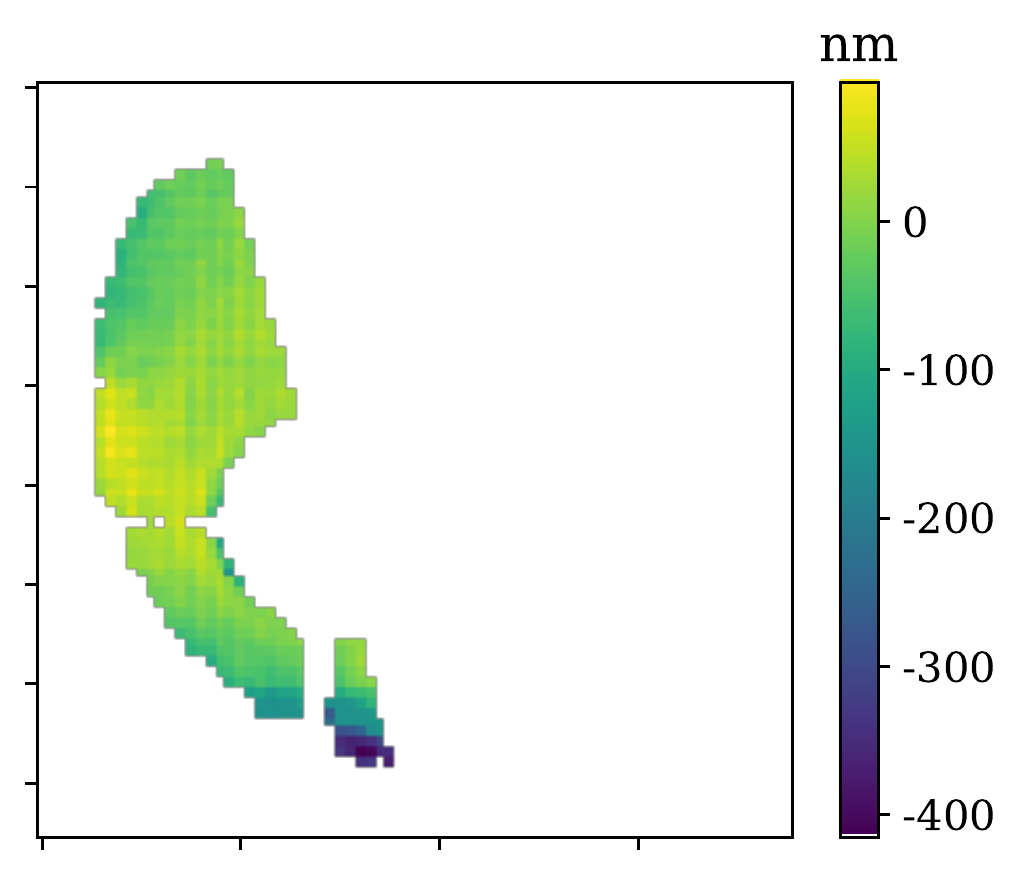}
      \caption{The mean unwrapped phase after tip, tilt, and  piston terms were subtracted  from each measurement for pixels where the \gls{NPZT} mirror moved more than $\pi/4$.}  \label{fig:lsq_avg_phase}
          \end{subfigure}
                             \hfill          
  \begin{subfigure}[t]{0.45\linewidth}
     \includegraphics[width=.9\textwidth]{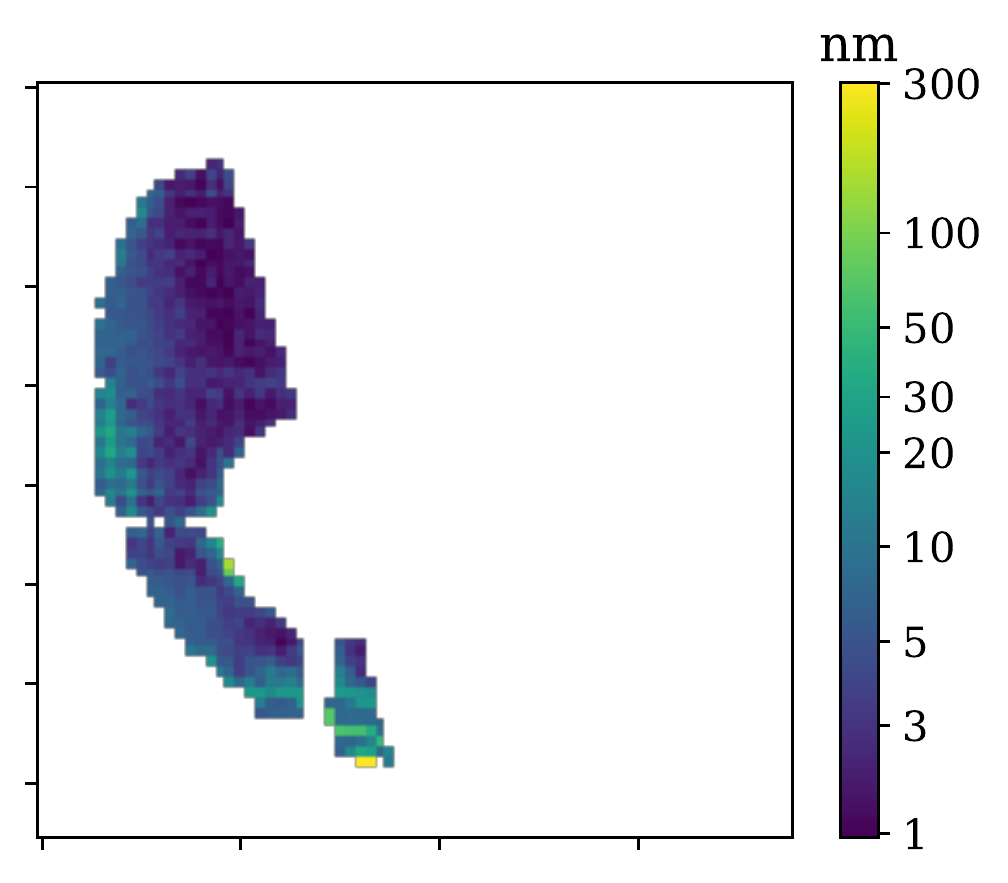}
 \caption{The \gls{WFS} phase sensing stability, the standard deviation of the measurement-to-measurement phase for pixels where the \gls{NPZT} mirror stepped more than $\pi/4$.}
    \label{fig:lsq_stability}
  \end{subfigure}
     \caption{\gls{WFS} pupil phase and error maps showing the phase was poorly measured where the \gls{NPZT} stepsize was smaller than $\pi/4$. 
   Across the region of the pupil where the steps were greater than $\pi/4$ the measured wavefront was flat, with stability of a few nanometers.
     Units are nanometers of wavefront error, calculated by assuming the phase error measured in radians is at 675 nm, the central wavelength of the \gls{VNC}.
}
\end{figure}

\section{Wavefront Sensor Results}\label{sec:results}
This section presents a stability analysis of the in-flight wavefront error during the initial period of consecutive coarse-mode measurements (Frames 28-37) recorded after the \gls{FPS} first locked and before repointing was attempted. 
These ten ABCD image sets and calculated phase measurements correspond to the best instrument focus and the highest visibility fringes.
Fig. \ref{fig:lsq_avg_uncertainty} shows the mean precision of each phase measurement, the phase  fitting error including photon noise, for each of the pixels across the pupil.    
This mean measurement error across the interfering pupil becomes highly uncertain (exceeding $\lambda/2$) where the phase steps become small (see the phase step map, Fig. \ref{fig:RailedStepMap}).

In order to compare the wavefront sensor precision to the expected performance of a fully stepping wavefront sensor, we define a sufficiently stepping region and exclude the regions of the pupil where the \gls{NPZT} step-size was relatively small (below $\pi/4$). 
This sufficiently stepped region has an area of 759 \gls{WFS} pixels, providing a relatively large sample with which to assess the instrument stability and sensing precision.    
 Fig. \ref{fig:lsq_avg_uncertainty_good} shows the mean fitting error across this region.

To assess the wavefront sensing stability, \gls{TTP} errors are subtracted from each phase measurement by fitting a 2D plane to the remaining phase pixels because the \gls{NPZT} was operating in a closed loop correction mode and each \gls{NPZT} position varied slightly.
Due to the shearing mechanism of the nuller, this also removes errors due to changes in telescope focus and astigmatism\cite{douglas_advancing_2016}.
Fig. \ref{fig:lsq_avg_phase} shows the mean of the \gls{TTP} subtracted phase measurements in the >$\pi/4$ step region.
Fig. \ref{fig:lsq_stability} shows a map of the stability as expressed by the measurement-to-measurement  standard deviation in the phase, $\Delta\phi'$, for each pixel.


\begin{figure}[htp]
\centering
       \includegraphics[width=0.65\textwidth]{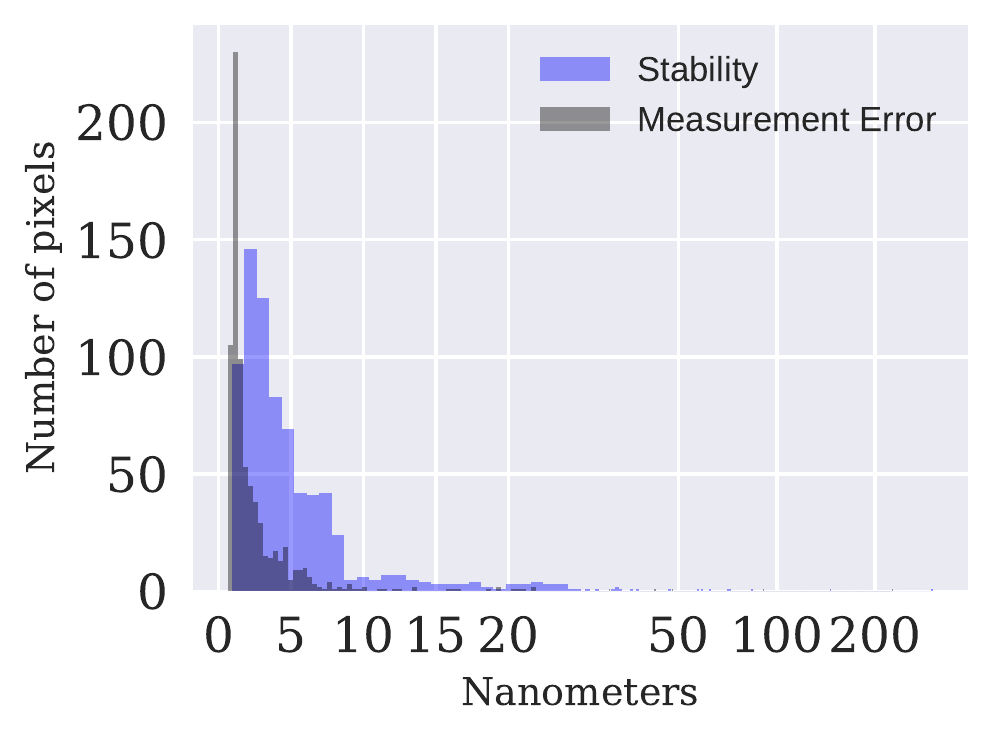}
          \caption{Histograms of the uncertainty in the wavefront sensor pixels where  $\delta$ exceeded $\pi/4$ for the ten high-visibility measurements after subtraction of a best-fit plane. 
   The higher-order stability between measurements shows the \gls{DM} and \gls{WFS} were both sufficiently stable to reach the expected contrasts, although the per-pixel stability is significantly higher than the fitting error.}
   \label{fig:wfs_sigma_hist}
\end{figure}

Histograms in Fig. \ref{fig:wfs_sigma_hist} show the distribution of uncertainty for both the mean measurement error and the stability for the sufficient step region.
The mean phase stability in this region is 6.7 nm/pixel and the standard deviation is 17.6  nm/pixel, for all the pixels in the included measurement-to-measurement stability map, Fig. \ref{fig:lsq_stability}.
 The distribution of stability per pixel, Fig. \ref{fig:wfs_sigma_hist}, exhibits a long tail, or a few pixels with very high uncertainty, which corresponds to the low visibility region at the bottom edge of the pupil. 
  Sigma-clipping the sufficient step pixels, removing the outliers iteratively beyond 5$\sigma$, leaves 741 pixels with a mean stability of  4.8 $\pm$ 4.1 nm.
The mean measurement error is 3.0  nm/pixel and the standard deviation is 9.9 nm/pixel, measured by taking the mean  and standard deviation of the sufficient step fitting error map, Fig. \ref{fig:lsq_avg_uncertainty_good}.   
Sigma-clipping leaves 737 pixels with a mean measurement error of  2.1 $\pm$ 1.7 nm.


\begin{figure}[htp]
  \centering  
\begin{subfigure}[b]{0.55\linewidth}
           \includegraphics[width=\textwidth]{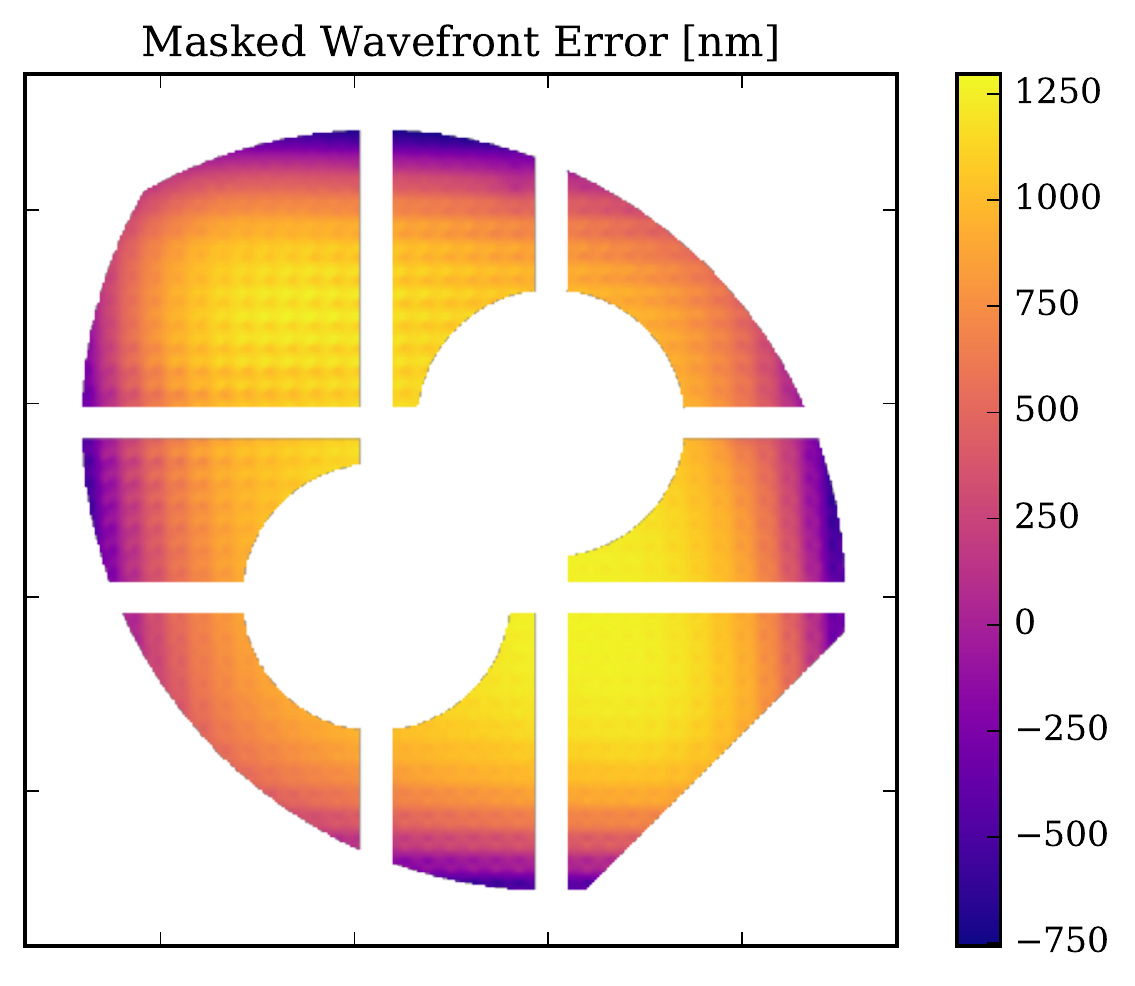}
  \caption{Unactuated wavefront error map (twice the surface error) of a flight Kilo-DM mirror, digitally masked to match the \gls{VNC} Lyot mask. 
  Raw surface measurement supplied by Boston Micromachines Corporation. }\label{fig:dm_off_meas}
       \end{subfigure}
                        \hfill          
               \hfill       
                         \begin{subfigure}[b]{0.4\linewidth}
           \includegraphics[width=\linewidth]{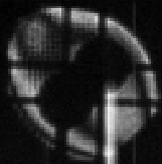}
  \caption{Unactuated \gls{DM} fringe pattern in the laboratory through \gls{VNC}, showing more than a wave of curvature without the flat map.
   The light vertical bar is unsubtracted dark noise. }\label{fig:dm_off_fringes}
       \end{subfigure}
        \caption{Laboratory measurements showing  the unpowered \gls{DM} surface curvature due to residual manufacturing stresses which is not present in flight data, indicating the \gls{DM} actuated appropriately in space. }
   \label{fig:DMflatMap}
\end{figure}
 
\subsection{Discussion}\label{sec:discussion}
When compared to the unpowered \gls{DM} surface map, Fig. \ref{fig:dm_off_meas}, and the corresponding fringe pattern, Fig. \ref{fig:dm_off_fringes}, the   in-flight wavefront error lacks  the stress-induced surface concavity observed when the  \gls{DM} is unpowered.
This indicates that the \gls{DM} was powered on, and the measured actuators were responding appropriately, moving  to  the  commanded default positions.

The observed fringe visibility provides a first order estimate of the \gls{NPZT} mirror position relative to the center of the fringe packet. 
The translating mirror was originally aligned to the mid-point of the 8.5 micron range.
As discussed in Sec. \ref{subsec:phasecalc}, the coherence length is approximately 3 $\mu$m. 
The visibility was moderate, with approximately one wave of phase tilt across the pupil  in the first measurement (Fig. \ref{fig:ABCD28}). 
Thus, the central fringe was within a few microns of the \gls{NPZT} limit, implying a total motion of no more than 7 $\mu$m from the original alignment at the center of the \gls{NPZT} range. 
Post-flight laboratory testing found a  displacement consistent with this estimate; the optimal \gls{NPZT} mirror position has shifted several microns compared to the prelaunch alignment, implying minimal movement during re-entry and recovery. 

This shift is likely due to motion of the 6-degrees-of-freedom \gls{DM} mount. 
Whether this shift occurred due to launch forces or upon reentry and impact cannot be definitively determined since a flight shift of the \gls{NPZT} or \gls{DM} mounts could also have been due to a large temperature gradient within the payload.
However, the \gls{DM} mount temperature was stable to approximately 0.3$^\circ$ C in flight\cite{douglas_advancing_2016} and prior laboratory tests found the \gls{VNC} path length has an approximately 700 nm/$^\circ$C dependency\cite{mendillo_scattering_2013}, well within the range of the \gls{NPZT} to correct.
 Random vibration is a large contributor to optical bench instability in spacecraft\cite{edeson_dimensional_2009}, making the sounding rocket launch environment the most likely cause of a few micron displacement. 
This suggests improved mount designs, or additional active correction stages, are required for future missions with micron-scale alignment tolerances.
While the payload underwent random vibration before launch, the launch of a sounding rocket also subjects the payload to a large continuous acceleration, which is difficult to replicate in testing and may have contributed inelastic deformation or slippage of the \gls{DM} mounting structure.

The measurement-to-measurement stability of the wavefront sensor measurements indicates the deformable mirror surface and the \gls{NPZT} $\delta$ step positions were relatively stable and within a factor of two of the laboratory measured stability for the \gls{VNC} alone. 
Four wavefront sensor pixels sample each \gls{DM} actuator; thus, assuming the actuator errors are uncorrelated, the uncertainty in phase per actuator is half the values reported herein.
The measurement error in the sufficiently stepping region reached the expected 2 nm wavefront error floor\cite{mendillo_scattering_2013} due to photon noise for Rigel.
Even including instability, had these \gls{WFS} measurements been applied to the \gls{DM} to correct phase mismatch between the nuller arms, Eq. \ref{eq:phase_leakage} shows that the total leakage per actuator due to residual phase error, neglecting other sources of leakage, would have been approximately $1\times10^{-4}L_\star$.

\section{Conclusions}\label{sec:conclusion}
The \gls{PICTURE} sounding rocket program has advanced exoplanet imaging technology by translating laboratory demonstrated concepts into deployed spaceflight hardware. 
The PICTURE program has previously demonstrated an \gls{FPS} that provides precision pointing, and this analysis shows active wavefront sensing precision at nanometer scales with a \gls{VNC}.
The second PICTURE flight also marks the first operation and measurement of a deformable mirror for high-contrast imaging in space. 
Several upcoming missions will continue  progress in high-contrast imaging from space.
The  Planetary Imaging Concept Testbed Using a Recoverable Experiment - Coronagraph (PICTURE -- C) high-altitude balloon will demonstrate wavefront sensing and control of both phase and amplitude with a vector vortex coronagraph\cite{cook_planetary_2015}. 
The Deformable Mirror Demonstration Mission CubeSat\cite{cahoy_wavefront_2013} is being built as a \gls{MEMS} deformable mirror testbed in low-Earth orbit.
  The  \gls{WFIRST} coronagraph instrument\cite{spergel_wide-field_2015} is planned to demonstrate wavefront sensing and control for both shaped pupil and hybrid-lyot internal coronagraphs behind an obscured telescope aperture during a multiyear mission.

\acknowledgments     

 The PICTURE-B team would like to thank the NASA Sounding Rocket Program Office, the Wallops Flight Facility, and the Orbital ATK  NSROC II team for their support, particularly our mission managers: Christine Chamberlain and David Jennings. 
We are also deeply indebted to everyone in the \gls{wsmr} Naval Research Rocket Support Office and NMSU Physical Science Laboratory teams for their leadership and assistance.
 
This work was funded by NASA grants NNG05WC17G, NNX11AD53G, NNX13AD50G, NNX15AG23G, and through graduate fellowships awarded to E.S. Douglas by the Massachusetts Space Grant Consortium.   
Computing resources were provided by two interfaces to the Massachusetts Green Computing Facility \cite{brown_green_2012},  MIT Research Computing, and the Boston University Scientific Computing Cluster. 

Special thanks to Brian A. Hicks of NASA Goddard Space Flight Facility, Benjamin F. Lane of MIT Draper Laboratory, and Shanti Rao and J. Kent Wallace of the Jet Propulsion Laboratory. 
The Boston University Scientific Instrument Facility worked tirelessly to support to integration of both PICTURE  payloads. 
 Paul Bierden, Charles Conway, and the rest of the staff of Boston Micromachines Corporation provided invaluable support to this project.  
 The staff at AOA Xinetics Northrop Grumman and John G. Daly of Vector Engineering provided essential support to the refurbishment of the flight telescope.
E.S.D. would like especially thank Catherine Espaillat, Alan Marscher, Donald W. McCarthy, and Michael Mendillo for their valuable input.
 This research made use of community-developed core Python packages, including: Astropy\cite{the_astropy_collaboration_astropy:_2013}, Matplotlib\cite{hunter_matplotlib:_2007}, SciPy\cite{jones_scipy:_2001}, and
the IPython Interactive Computing architecture \cite{perez_ipython:_2007}.
Additional data analyses were done using IDL  (Exelis Visual Information Solutions, Boulder, Colorado).  
This research has made use of the SIMBAD database, operated at CDS, Strasbourg, France.

 \section*{Disclosures}     
The authors have no financial interests to disclose. 

\bibliography{../../../../MyLibrary}   
\bibliographystyle{spiebib}   

\newpage

\end{document}

%% file: acronyms.tex
\newacronym{AU}{AU}{Astronomical Unit [1.5e11 m]}  
\newacronym{pc}{pc}{ parsec }
\newacronym{mas}{mas}{ milliarcsecond }
\newacronym{nm}{nm}{ Nanometer  }
\newacronym{CTE}{CTE}{Coefficient of Thermal Expansion}

\newacronym{smc}{SMC}{Small Magellanic Cloud}
\newacronym{lmc}{LMC}{Large Magellanic Cloud}
\newacronym{ism}{ISM}{interstellar medium}
\newacronym{mw}{MW}{Milky Way}
\newacronym{epseri}{$\epsilon$ Eri}{Epsilon Eridani}

\newacronym{CFR}{CFR}{Complete Frequency Redistribution}

\newacronym{nasa}{NASA}{National Aeronautics and Space Agency}
\newacronym{esa}{ESA}{European Space Agency}
\newacronym{omi}{OMI}{\textit{Optical Mechanics Inc.}}
\newacronym{gsfc}{GSFC}{\gls{nasa} Goddard Space Flight Center}
\newacronym{stsci}{STScI}{Space Telescope Science Institute}
\newacronym{nsroc}{NSROC}{\gls{nasa} Sounding Rocket Operations Contract}
\newacronym{wff}{WFF}{\gls{nasa} Wallops Flight Facility}
\newacronym{wsmr}{WSMR}{White Sands Missile Range}

\newacronym{irac}{IRAC}{Infrared Array Camera}
\newacronym[plural=CCDs, firstplural=charge-coupled devices (CCDs)]{ccd}{CCD}{charge-coupled device}
\newacronym{DM}{DM}{Deformable Mirror}
\newacronym{MCP}{MCP}{ Microchannel Plate }
\newacronym{ipc}{IPC}{Image Proportional Counter}
\newacronym{cots}{COTS}{Commercial Off-The-Shelf}
\newacronym{ISR}{ISR}{Incoherent Scatter Radar }
\newacronym{atcamera}{AT}{Angle Tracker}
\newacronym{MEMS}{MEMS}{microelectromechanical systems}
\newacronym{QE}{QE}{quantum efficiency}
\newacronym{RTD}{RTD}{Resistance Temperature Detector}
\newacronym{PID}{PID}{Proportional-Integral-Derivative}

\newacronym{FOV}{FOV}{field-of-view}
\newacronym{NIR}{NIR}{near-infrared}
\newacronym{PV}{PV}{Peak-to-Valley}
\newacronym{MRF}{MRF}{Magnetorheological finishing}
\newacronym{AO}{AO}{Adaptive Optics}
\newacronym{TTP}{TTP}{tip, tilt, and piston}
\newacronym{FPS}{FPS}{fine pointing system}

\newacronym{acs}{ACS}{Attitude Control System}
\newacronym{orsa}{ORSA}{Ogive Recovery System Assembly}
\newacronym{gse}{GSE}{Ground Station Equipment}
\newacronym{FSM}{FSM}{Fast Steering Mirror}

\newacronym{WFS}{WFS}{wavefront sensor}
\newacronym{LSI}{LSI}{Lateral Shearing Interferometer}
\newacronym{VVC}{VVC}{Vector Vortex Coronagraph}
\newacronym{VNC}{VNC}{Visible Nulling Coronagraph}
\newacronym{CGI}{CGI}{Coronagraph Instrument}
\newacronym{IWA}{IWA}{Inner Working Angle}
\newacronym{OWA}{OWA}{Outer Working Angle}
\newacronym{NPZT}{N-PZT}{Nuller piezoelectric transducer}
\newacronym{OPD}{OPD}{Optical Path Difference}
\newacronym{WFCS}{WFCS}{Wavefront Control System}

\newacronym{HST}{HST}{ Hubble Space Telescope}
\newacronym{GPS}{GPS}{Global Positioning System}
\newacronym{ISS}{ISS}{International Space Station}
\newacronym[description=Advanced CCD Imaging Spectrometer]{acis}{ACIS}{Advanced \gls{ccd} Imaging Spectrometer}
\newacronym{stis}{STIS}{\textit{Space Telescope Imaging Spectrograph}}
\newacronym{mcp}{MCP}{Microchannel Plate}
\newacronym{jwst}{JWST}{$\textit{James Webb Space Telescope}$}
\newacronym{fuse}{FUSE}{$\textit{FUSE}$}
\newacronym{galex}{GALEX}{$\textit{Galaxy Evolution Explorer}$}
\newacronym{spitzer}{Spitzer}{$\textit{Spitzer Space Telescope}$}
\newacronym{mips}{MIPS}{Multiband Imaging Photometer for \gls{spitzer}}
\newacronym{gissmo}{GISSMO}{Gas Ionization Solar Spectral Monitor}
\newacronym{iue}{IUE}{International Ultraviolet Explorer}
\newacronym{spinr}{SPINR}{$\textit{Spectrograph for Photometric Imaging with Numeric Reconstruction}$}
\newacronym{imager}{IMAGER}{$\textit{Interstellar Medium Absorption Gradient Experiment Rocket}$}
\newacronym{TPF-C}{TPF-C}{Terrestrial Planet Finder Coronagraph}
\newacronym{RAIDS}{RAIDS}{Atmospheric and Ionospheric Detection System }
\newacronym{mama}{MAMA}{Multi-Anode Microchannel Array}
\newacronym{ATLAST}{ATLAST}{Advanced Technology Large Aperture Space Telescope}
\newacronym{PICTURE}{PICTURE}{Planet Imaging Concept Testbed Using a Rocket Experiment}
\newacronym{LITES}{LITES}{Limb-imaging Ionospheric and Thermospheric
Extreme-ultraviolet Spectrograph}
\newacronym{LBT}{LBT}{Large Binocular Telescope}
\newacronym{LBTI}{LBTI}{Large Binocular Telescope Interferometer}
\newacronym{KIN}{KIN}{Keck Interferometer Nuller}
\newacronym{SHARPI}{SHARPI}{Solar High-Angular Resolution Photometric Imager}
\newacronym{IRAS}{IRAS}{Infrared Astronomical Satellite}
\newacronym{HARPS}{HARPS}{High Accuracy Radial velocity Planetary}
\newacronym{hstSTIS}{STIS}{Space Telescope Imaging Spectrograph}
\newacronym{spitzerIRAC}{IRAC}{Infrared Array Camera}
\newacronym{spitzerMIPS}{MIPS}{Multiband Imaging Photometer for Spitzer}
\newacronym{spitzerIRS}{IRS}{Infrared Spectrograph}
\newacronym{CHARA}{CHARA}{Center for High Angular Resolution Astronomy}
\newacronym{wfirst-afta}{WFIRST-AFTA}{Wide-Field InfrarRed Survey
Telescope-Astrophysics Focused Telescope Assets}
\newacronym{GPI}{GPI}{Gemini Planet Imager}
\newacronym{WFIRST}{WFIRST}{Wide-Field InfrarRed Survey Telescope}

\newacronym{AURIC}{AURIC}{The Atmospheric Ultraviolet Radiance Integrated Code} 
\newacronym{FFT}{FFT}{Fast Fourier Transform  }
\newacronym{MODTRAN}{MODTRAN   }{ MODerate resolution atmospheric TRANsmission }
\newacronym{idl}{IDL}{$\textit {Interactive Data Language}$}
\newacronym[sort=NED,description=NASA/IPAC Extragalactic Database]{ned}{NED}{\gls{nasa}/\gls{ipac} Extragalactic Database}
\newacronym{iraf}{IRAF}{Image Reduction and Analysis Facility}
\newacronym{wcs}{WCS}{World Coordinate System}
\newacronym{pegase}{PEGASE}{$\textit{Projet d'Etude des GAlaxies par Synthese Evolutive}$}
\newacronym{dirty}{DIRTY}{$\textit{DustI Radiative Transfer, Yeah!}$}
\newacronym{MGHPCC}{MGHPCC}{Massachusetts Green High Performance
Computing Center}

\newacronym{MSIS}{MSIS}{Mass Spectrometer Incoherent Scatter Radar}
\newacronym{nmf2}{$N_m$}{F2-Region Peak density}
\newacronym{hmf2}{$h_m$}{F2-Region Peak height}
\newacronym{H}{$H$}{F2-Region Scale Height}

\newacronym{isr}{ISR}{Incoherent Scatter Radar}
\newacronym[description=TLA Within Another Acronym]{twaa}{TWAA}{\gls{tla} Within Another Acronym}
\newacronym[plural=SNe, firstplural=Supernovae (SNe)]{sn}{SN}{Supernova}
\newacronym{EUV}{EUV}{Extreme-Ultraviolet }
\newacronym{EUVS}{EUVS}{\gls{EUV} Spectrograph}
\newacronym{F2}{F2}{Ionospheric Chapman F Layer }
\newacronym{F10.7}{F10.7}{ 10.7 cm radio flux [10$^{-22}$ W m$^{-2}$ Hz$^{-1}$]  }
\newacronym{FUV}{FUV}{ Far-Ultraviolet }
\newacronym{IR}{IR}{Infrared}
\newacronym{MUV}{MUV}{Mid-Ultraviolet }
\newacronym{NUV}{NUV}{Near-Ultraviolet }
\newacronym{O$^+$}{O$^+$}{Singly Ionized Oxygen  Atom }
\newacronym{OI}{OI}{Neutral Atomic Oxygen Spectroscopic State }
\newacronym{OII}{OII}{Singly Ionized Atomic Oxygen Spectroscopic State }
\newacronym{PSF}{PSF}{Point Spread Function}
\newacronym{$R_E$}{$R_E$}{ Earth Radii [$\approx$ 6400 km]  }
\newacronym{RV}{RV}{Radial Velocity}
\newacronym{UV}{UV}{Ultraviolet }
\newacronym{WFE}{WFE}{Wavefront Error}
\newacronym{sed}{SED}{Spectral Energy Distribution}
\newacronym{nir}{NIR}{near-infrared}
\newacronym{mir}{MIR}{mid-infrared}
\newacronym{ir}{IR}{infrared}
\newacronym{uv}{UV}{ultraviolet}
\newacronym[plural=PAHs, firstplural=Polycyclic Aromatic Hydrocarbons (PAHs)]{pah}{PAH}{Polycyclic Aromatic Hydrocarbon}
\newacronym{obsid}{OBSID}{Observation Identification}
\newacronym{SZA}{SZA}{Solar Zenith Angle}
\newacronym{TLE}{TLE}{Two Line Element set}
\newacronym{DOF}{DOF}{degrees-of-freedom}
\newacronym{PZT}{PZT}{lead zirconate titanate}

\newacronym{PCA}{PCA}{Principal Component Analysis}
\newacronym{fwhm}{FWHM}{Full-Width-Half Maximum}
\newacronym{RMS}{RMS}{root mean squared}
\newacronym{RMSE}{RMSE}{root mean squared error}
\newacronym{MCMC}{MCMC}{Marcov chain Monte Carlo}
\newacronym{DIT}{DIT}{Discrete Inverse Theory}
\newacronym{SNR}{SNR}{signal-to-noise ratio}
\newacronym{PSD}{PSD}{Power Spectral Density}

%% file: 2017july1_local_figs.bbl
\begin{thebibliography}{10}

\bibitem{turnbull_spectrum_2006}
M.~C. Turnbull, W.~A. Traub, K.~W. Jucks, N.~J. Woolf, M.~R. Meyer, N.~Gorlova,
  M.~F. Skrutskie, and J.~C. Wilson, ``Spectrum of a habitable world:
  {Earthshine} in the near-infrared,'' {\em The Astrophysical Journal}~{\bf
  644}(1), p.~551, 2006.

\bibitem{cahoy_exoplanet_2010}
K.~L. Cahoy, M.~S. Marley, and J.~J. Fortney, ``Exoplanet {Albedo} {Spectra}
  and {Colors} as a {Function} of {Planet} {Phase}, {Separation}, and
  {Metallicity},'' {\em ApJ}~{\bf 724}, pp.~189--214, Nov. 2010.

\bibitem{spitzer_beginnings_1962}
L.~Spitzer, ``The beginnings and future of space astronomy,'' {\em American
  Scientist}~{\bf 50}(3), pp.~473--484, 1962.

\bibitem{seager_exo-s:_2015}
S.~Seager, W.~Cash, S.~Domagal-Goldman, and {others}, ``Exo-{S}: starshade
  probe-class exoplanet direct imaging mission concept {Final} {Report},'' {\em
  available at exep. jpl. nasa. gov/stdt} , 2015.

\bibitem{traub_direct_2010}
W.~A. Traub and B.~R. Oppenheimer, ``Direct imaging of exoplanets,'' in {\em
  Exoplanets},  pp.~111--156, University of Arizona Press, Tucson, AZ, USA,
  2010.
\newblock Seager, S., ed.

\bibitem{guyon_theoretical_2006}
O.~Guyon, E.~A. Pluzhnik, M.~J. Kuchner, B.~Collins, and S.~T. Ridgway,
  ``Theoretical {Limits} on {Extrasolar} {Terrestrial} {Planet} {Detection}
  with {Coronagraphs},'' {\em The Astrophysical Journal Supplement Series}~{\bf
  167}, pp.~81--99, Nov. 2006.

\bibitem{trauger_laboratory_2007}
J.~T. Trauger and W.~A. Traub, ``A laboratory demonstration of the capability
  to image an {Earth}-like extrasolar planet,'' {\em Nature}~{\bf 446},
  pp.~771--773, Apr. 2007.

\bibitem{roberge_exozodiacal_2012}
A.~Roberge, C.~H. Chen, R.~Millan-Gabet, A.~J. Weinberger, P.~M. Hinz, K.~R.
  Stapelfeldt, O.~Absil, M.~J. Kuchner, and G.~Bryden, ``The {Exozodiacal}
  {Dust} {Problem} for {Direct} {Observations} of {Exo}-{Earths},'' {\em
  Publications of the Astronomical Society of the Pacific}~{\bf 124},
  pp.~799--808, Aug. 2012.

\bibitem{turnbull_search_2012}
M.~C. Turnbull, T.~Glassman, A.~Roberge, W.~Cash, C.~Noecker, A.~Lo, B.~Mason,
  P.~Oakley, and J.~Bally, ``The {Search} for {Habitable} {Worlds}. 1. {The}
  {Viability} of a {Starshade} {Mission},'' {\em Publications of the
  Astronomical Society of the Pacific}~{\bf 124}(915), p.~418, 2012.

\bibitem{stark_detectability_2008}
C.~C. Stark and M.~J. Kuchner, ``The {Detectability} of {Exo}‐{Earths} and
  {Super}‐{Earths} {Via} {Resonant} {Signatures} in {Exozodiacal} {Clouds},''
  {\em ApJ}~{\bf 686}, pp.~637--648, Oct. 2008.

\bibitem{defrere_nulling_2010}
D.~Defrère, O.~Absil, R.~den Hartog, C.~Hanot, and C.~Stark, ``Nulling
  interferometry: impact of exozodiacal clouds on the performance of future
  life-finding space missions,'' {\em Astronomy and Astrophysics}~{\bf 509},
  p.~9, Jan. 2010.

\bibitem{hinz_large_2003}
P.~M. Hinz, J.~R.~P. Angel, J.~McCarthy, Donald~W., W.~F. Hoffman, and C.~Y.
  Peng, ``The {Large} {Binocular} {Telescope} interferometer,'' in {\em Proc.
  {SPIE}},   {\bf 4838}, pp.~108--112, 2003.

\bibitem{eiroa_dust_2013}
C.~Eiroa, J.~P. Marshall, A.~Mora, B.~Montesinos, O.~Absil, J.~C. Augereau,
  A.~Bayo, G.~Bryden, W.~Danchi, C.~del Burgo, S.~Ertel, M.~Fridlund, A.~M.
  Heras, A.~V. Krivov, R.~Launhardt, R.~Liseau, T.~Löhne, J.~Maldonado, G.~L.
  Pilbratt, A.~Roberge, J.~Rodmann, J.~Sanz-Forcada, E.~Solano, K.~Stapelfeldt,
  P.~Thébault, S.~Wolf, D.~Ardila, M.~Arévalo, C.~Beichmann, V.~Faramaz,
  B.~M. González-García, R.~Gutiérrez, J.~Lebreton, R.~Martínez-Arnáiz,
  G.~Meeus, D.~Montes, G.~Olofsson, K.~Y.~L. Su, G.~J. White, D.~Barrado,
  M.~Fukagawa, E.~Grün, I.~Kamp, R.~Lorente, A.~Morbidelli, S.~Müller,
  H.~Mutschke, T.~Nakagawa, I.~Ribas, and H.~Walker, ``{DUst} {Around} {NEarby}
  {Stars}. {The} survey observational results,'' {\em Astronomy \&
  Astrophysics}~{\bf 555}, p.~A11, July 2013.
\newblock arXiv: 1305.0155.

\bibitem{shao_nulling_2006}
M.~Shao, B.~M. Levine, J.~K. Wallace, G.~S. Orton, E.~Schmidtlin, B.~F. Lane,
  S.~Seager, V.~Tolls, R.~G. Lyon, R.~Samuele, D.~J. Tenerelli, R.~Woodruff,
  and J.~Ge, ``A nulling coronagraph for {TPF}-{C},'' {\em Proc. SPIE} ,
  pp.~626517--626517, June 2006.

\bibitem{samuele_experimental_2007}
R.~Samuele, J.~Wallace, E.~Schmidtlin, M.~Shao, B.~Levine, and S.~Fregoso,
  ``Experimental {Progress} and {Results} of a {Visible} {Nulling}
  {Coronagraph},'' in {\em 2007 {IEEE} {Aero}. {Conf}.},  pp.~1 --7, Mar. 2007.

\bibitem{rao_path_2008}
S.~R. Rao, J.~K. Wallace, R.~Samuele, S.~Chakrabarti, T.~Cook, B.~Hicks,
  P.~Jung, B.~Lane, B.~M. Levine, C.~Mendillo, E.~Schmidtlin, M.~Shao, and
  J.~B. Stewart, ``Path length control in a nulling coronagraph with a {MEMS}
  deformable mirror and a calibration interferometer,'' in {\em Proc. {SPIE}},
   {\bf 6888}, pp.~68880B--68880B, Feb. 2008.

\bibitem{mendillo_flight_2012}
C.~B. Mendillo, S.~Chakrabarti, T.~A. Cook, B.~A. Hicks, and B.~F. Lane,
  ``Flight demonstration of a milliarcsecond pointing system for direct
  exoplanet imaging,'' {\em Appl. Opt.}~{\bf 51}, pp.~7069--7079, Oct. 2012.

\bibitem{mendillo_picture:_2012}
C.~B. Mendillo, B.~A. Hicks, T.~A. Cook, T.~G. Bifano, D.~A. Content, B.~F.
  Lane, B.~M. Levine, D.~Rabin, S.~R. Rao, R.~Samuele, E.~Schmidtlin, M.~Shao,
  J.~K. Wallace, and S.~Chakrabarti, ``{PICTURE}: a sounding rocket experiment
  for direct imaging of an extrasolar planetary environment,'' in {\em Proc.
  {SPIE}},   {\bf 8442}, Sept. 2012.

\bibitem{douglas_end--end_2015}
E.~S. Douglas, K.~Hewasawam, C.~B. Mendillo, K.~L. Cahoy, T.~A. Cook, S.~C.
  Finn, G.~A. Howe, M.~J. Kuchner, N.~K. Lewis, A.~D. Marinan, D.~Mawet, and
  S.~Chakrabarti, ``End-to-end simulation of high-contrast imaging systems:
  methods and results for the {PICTURE} mission family,'' in {\em Proc.
  {SPIE}},   {\bf 9605}, pp.~96051A--96051A--13, 2015.

\bibitem{chakrabarti_planet_2016}
S.~Chakrabarti, C.~B. Mendillo, T.~A. Cook, J.~F. Martel, S.~C. Finn, G.~A.
  Howe, K.~Hewawasam, and E.~S. Douglas, ``Planet {Imaging} {Coronagraphic}
  {Technology} {Using} a {Reconfigurable} {Experimental} {Base}
  ({PICTURE}-{B}): {The} {Second} in the {Series} of {Suborbital} {Exoplanet}
  {Experiments},'' {\em Journal of Astronomical Instrumentation}~{\bf 05},
  p.~1640004, Mar. 2016.

\bibitem{di_folco_near-infrared_2007}
E.~di~Folco, O.~Absil, J.-C. Augereau, A.~Mérand, V.~Coudé~du Foresto,
  F.~Thévenin, D.~Defrère, P.~Kervella, T.~A. ten Brummelaar, H.~A.
  McAlister, S.~T. Ridgway, J.~Sturmann, L.~Sturmann, and N.~H. Turner, ``A
  near-infrared interferometric survey of debris disk stars. {I}. {Probing} the
  hot dust content around ɛ {Eridani} and τ {Ceti} with {CHARA}/{FLUOR},''
  {\em A\&A}~{\bf 475}, pp.~243--250, Nov. 2007.

\bibitem{backman_epsilon_2009}
D.~Backman, M.~Marengo, K.~Stapelfeldt, K.~Su, D.~Wilner, C.~D. Dowell,
  D.~Watson, J.~Stansberry, G.~Rieke, T.~Megeath, G.~Fazio, and M.~Werner,
  ``Epsilon {Eridani}'s {Planetary} {Debris} {Disk}: {Structure} and {Dynamics}
  {Based} on {Spitzer} and {Caltech} {Submillimeter} {Observatory}
  {Observations},'' {\em ApJ}~{\bf 690}, pp.~1522--1538, Jan. 2009.

\bibitem{reidemeister_cold_2011}
M.~Reidemeister, A.~V. Krivov, C.~C. Stark, J.-C. Augereau, T.~Löhne, and
  S.~Müller, ``The cold origin of the warm dust around \textit{ε}
  {Eridani},'' {\em A\&A}~{\bf 527}, p.~A57, Jan. 2011.

\bibitem{bracewell_detecting_1978}
R.~N. Bracewell, ``Detecting nonsolar planets by spinning infrared
  interferometer,'' {\em Nature}~{\bf 274}, pp.~780--781, Aug. 1978.

\bibitem{serabyn_nulling_2000}
E.~Serabyn, ``Nulling interferometry: symmetry requirements and experimental
  results,'' in {\em Proc. {SPIE}},   {\bf 4006}, pp.~328--339, 2000.

\bibitem{hicks_high-contrast_2014}
B.~A. Hicks, R.~G. Lyon, M.~R. Bolcar, M.~Clampin, and P.~Petrone,
  ``High-contrast visible nulling coronagraph for segmented and arbitrary
  telescope apertures,''  {\bf 9143}, pp.~91432S--91432S--11, 2014.

\bibitem{beichman_status_2006}
C.~Beichman, P.~Lawson, O.~Lay, A.~Ahmed, S.~Unwin, and K.~Johnston, ``Status
  of the terrestrial planet finder interferometer ({TPF}-{I}),''  {\bf 6268},
  pp.~62680S--62680S--9, 2006.

\bibitem{shao_visible_2004}
M.~Shao, J.~K. Wallace, B.~M. Levine, and D.~T. Liu, ``Visible nulling
  interferometer,'' in {\em Proc. {SPIE}},   {\bf 5487}, pp.~1296--1303, Oct.
  2004.

\bibitem{shao_calibration_2005}
M.~Shao, J.~J. Green, B.~Lane, J.~K. Wallace, B.~M. Levine, R.~Samuele, S.~Rao,
  and E.~Schmidtlin, ``Calibration of {Residual} {Speckle} {Pattern} in a
  {Coronagraph},'' {\em Proceedings of the International Astronomical
  Union}~{\bf 1}(Colloquium C200), pp.~525--528, 2005.

\bibitem{lyon_visible_2006}
R.~G. Lyon, M.~Clampin, R.~Woodruff, G.~Vasudevan, M.~Shao, M.~Levine,
  G.~Melnick, V.~Tolls, P.~Petrone, and P.~Dogoda, ``Visible nulling
  coronagraphy for exo-planetary detection and characterization,'' {\em IAU
  Colloquim}~{\bf 200}, 2006.

\bibitem{levine_visible_2006}
B.~M. Levine, F.~Aguayo, T.~Bifano, S.~F. Fregoso, J.~J. Green, B.~F. Lane,
  D.~T. Liu, B.~Mennesson, S.~Rao, R.~Samuele, M.~Shao, E.~Schmidtlin,
  E.~Serabyn, J.~Stewart, and J.~K. Wallace, ``The visible nulling coronagraph:
  architecture definition and technology development status,'' in {\em Proc.
  {SPIE}},  J.~C. Mather, H.~A. MacEwen, and M.~W.~M. de~Graauw, eds.,
  pp.~62651A--62651A--13, June 2006.

\bibitem{morgan_nulling_2000}
R.~M. Morgan, J.~H. Burge, and N.~J. Woolf, ``Nulling interferometric beam
  combiner utilizing dielectric plates: experimental results in the visible
  broadband,'' in {\em Proc. {SPIE}},   {\bf 4006}, pp.~340--348, July 2000.

\bibitem{angel_imaging_1997}
J.~R.~P. Angel and N.~J. Woolf, ``An {Imaging} {Nulling} {Interferometer} to
  {Study} {Extrasolar} {Planets},'' {\em ApJ}~{\bf 475}, p.~373, Jan. 1997.

\bibitem{hicks_nulling_2012}
B.~A. Hicks, {\em Nulling interferometers for space-based high contrast visible
  imaging and measurement of exoplanetary environments}.
\newblock PhD thesis, Boston University, Boston, MA, USA, 2012.

\bibitem{douglas_advancing_2016}
E.~S. Douglas, {\em Advancing spaceborne tools for the characterization of
  planetary ionospheres and circumstellar environments}.
\newblock PhD thesis, Boston University, Boston, MA, USA, 2016.

\bibitem{krist_phase-retrieval_1995}
J.~E. Krist and C.~J. Burrows, ``Phase-retrieval analysis of pre- and
  post-repair {Hubble} {Space} {Telescope} images,'' {\em Applied Optics}~{\bf
  34}, p.~4951, Aug. 1995.

\bibitem{vosteen_wavefront_2009}
L.~L.~A. Vosteen, F.~Draaisma, W.~P. Van~Werkhoven, L.~J.~M. Van~Riel, M.~H.
  Mol, and G.~den Ouden, ``Wavefront sensor for the {ESA}-{GAIA} mission,'' in
  {\em Astronomical and {Space} {Optical} {Systems}], {Society} of
  {Photo}-{Optical} {Instrumentation} {Engineers} ({SPIE}) {Conference}
  {Series}},   {\bf 7439}, p.~743914, 2009.

\bibitem{feinberg_trl-6_2007}
L.~D. Feinberg, B.~H. Dean, D.~L. Aronstein, C.~W. Bowers, W.~Hayden, R.~G.
  Lyon, R.~Shiri, J.~S. Smith, D.~S. Acton, L.~Carey, A.~Contos, E.~Sabatke,
  J.~Schwenker, D.~Shields, T.~Towell, F.~Shi, and L.~Meza, ``{TRL}-6 for
  {JWST} wavefront sensing and control,'' p.~668708, Sept. 2007.

\bibitem{greenbaum_-focus_2016}
A.~Z. Greenbaum and A.~Sivaramakrishnan, ``In-focus wavefront sensing using
  non-redundant mask-induced pupil diversity,'' {\em Opt. Express, OE}~{\bf
  24}, pp.~15506--15521, July 2016.

\bibitem{shi_low_2016}
F.~Shi, K.~Balasubramanian, R.~Bartos, R.~Hein, R.~Lam, M.~Mandic, D.~Moore,
  J.~Moore, K.~Patterson, I.~Poberezhskiy, J.~Shields, E.~Sidick, H.~Tang,
  T.~Truong, J.~K. Wallace, X.~Wang, and D.~W. Wilson, ``Low order wavefront
  sensing and control for {WFIRST} coronagraph,'' in {\em Proc. {SPIE}},
  p.~990418, July 2016.

\bibitem{wyant_use_1975}
J.~C. Wyant, ``Use of an ac heterodyne lateral shear interferometer with
  real-time wavefront correction systems,'' {\em Appl. Opt.}~{\bf 14},
  pp.~2622--2626, Nov. 1975.

\bibitem{ealey_xinetics_1994}
M.~A. Ealey and J.~A. Wellman, ``Xinetics low-cost deformable mirrors with
  actuator replacement cartridges,'' in {\em 1994 {Symposium} on {Astronomical}
  {Telescopes} \& {Instrumentation} for the 21st {Century}},  pp.~680--687,
  International Society for Optics and Photonics, 1994.

\bibitem{huang_experimental_2015}
L.~Huang, X.~Ma, M.~Gong, and Q.~Bian, ``Experimental investigation of the
  deformable mirror with bidirectional thermal actuators,'' {\em Optics
  Express}~{\bf 23}, p.~17520, June 2015.

\bibitem{lemmer_mathematical_2016}
A.~J. Lemmer, I.~M. Griffiths, T.~D. Groff, A.~W. Rousing, and N.~J. Kasdin,
  ``Mathematical and computational modeling of a ferrofluid deformable mirror
  for high-contrast imaging,''  {\bf 9912}, pp.~99122K--99122K--15, 2016.

\bibitem{bifano_microelectromechanical_1999}
T.~G. Bifano, J.~Perreault, R.~K. Mali, and M.~N. Horenstein,
  ``Microelectromechanical deformable mirrors,'' {\em IEEE Journal of Selected
  Topics in Quantum Electronics}~{\bf 5}, pp.~83--89, Jan. 1999.

\bibitem{morzinski_mems_2012}
K.~M. Morzinski, A.~P. Norton, J.~W. Evans, L.~Reza, S.~A. Severson, D.~Dillon,
  M.~Reinig, D.~T. Gavel, S.~Cornelissen, and B.~A. Macintosh, ``{MEMS}
  practice: from the lab to the telescope,'' in {\em {SPIE} {MOEMS}-{MEMS}},
  pp.~825304--825304, 2012.

\bibitem{yoo_mems_2009}
B.-W. Yoo, J.-H. Park, I.~H. Park, J.~Lee, M.~Kim, J.-Y. Jin, J.-A. Jeon, S.-W.
  Kim, and Y.-K. Kim, ``{MEMS} micromirror characterization in space
  environments,'' {\em Optics Express}~{\bf 17}, p.~3370, Mar. 2009.

\bibitem{fleming_calibration_2013}
B.~T. Fleming, S.~R. McCandliss, K.~Redwine, M.~E. Kaiser, J.~Kruk, P.~D.
  Feldman, A.~S. Kutyrev, M.~J. Li, S.~H. Moseley, O.~Siegmund, J.~Vallerga,
  and A.~Martin, ``Calibration and flight qualification of {FORTIS},''  {\bf
  8859}, pp.~88590Q--88590Q--12, 2013.

\bibitem{aguayo_fem_2014}
E.~J. Aguayo, R.~Lyon, M.~Helmbrecht, and S.~Khomusi, ``{FEM} correlation and
  shock analysis of a {VNC} {MEMS} mirror segment,'' in {\em {SPIE}
  {Astronomical} {Telescopes}+ {Instrumentation}},  pp.~91435C--91435C,
  International Society for Optics and Photonics, 2014.

\bibitem{bifano_micromachined_2000}
T.~G. Bifano, J.~A. Perreault, and P.~A. Bierden, ``Micromachined deformable
  mirror for optical wavefront compensation,'' in {\em International
  {Symposium} on {Optical} {Science} and {Technology}},  pp.~7--14,
  International Society for Optics and Photonics, 2000.

\bibitem{mendillo_scattering_2013}
C.~B. Mendillo, {\em Scattering properties of dust in {Orion} and the {Epsilon}
  {Eridani} exoplanetary system}.
\newblock PhD thesis, Boston University, Boston, MA, USA, 2013.

\bibitem{antonille_figure_2008}
S.~Antonille, D.~Content, D.~Rabin, S.~Wake, and T.~Wallace, ``Figure
  verification of a precision ultra-lightweight mirror: techniques and results
  from the {SHARPI}/{PICTURE} mirror at {NASA}/{GSFC},'' in {\em Proc. {SPIE}},
    {\bf 7011}, p.~28, Aug. 2008.

\bibitem{noauthor_sounding_2015}
{\em Sounding {Rocket} {Handbook}}, Sounding Rockets Program Office Suborbital
  \& Special Orbital Projects Directorate, Wallops Flight Facility Wallops
  Island, Virginia, USA. https://sites.wff.nasa.gov/code810/files/SRHB.pdf,
  2015.

\bibitem{yuhas_sounding_2012}
C.~Yuhas, ``Sounding {Rocket} {Program} {Update} to the {Heliophysics}
  {Subcommittee},'' July 2012.

\bibitem{bautz_progress_2004}
M.~W. Bautz, S.~E. Kissel, G.~Y. Prigozhin, B.~LaMarr, B.~E. Burke, and J.~A.
  Gregory, ``Progress in x-ray {CCD} sensor performance for the {Astro}-{E}2
  x-ray imaging spectrometer,'' in {\em Proc. {SPIE}},   {\bf 5501},
  pp.~111--122, 2004.

\bibitem{born_principles_1980}
M.~Born and E.~Wolf, {\em Principles of {Optics} {Electromagnetic} {Theory} of
  {Propagation}, {Interference} and {Diffraction} of {Light}}, Pergamon Press,
  6th~ed., 1980.

\bibitem{wyant_phase-shifting_2011}
J.~C. Wyant, ``Phase-{Shifting} {Interferometry},'' tech. rep.,
  https://wp.optics.arizona.edu/jcwyant/wp-content/uploads/sites/13/2016/08/Phase-Shifting-Interferometry.nb\_.pdf,
  2011.

\bibitem{branch_subspace_1999}
M.~Branch, T.~Coleman, and Y.~Li, ``A {Subspace}, {Interior}, and {Conjugate}
  {Gradient} {Method} for {Large}-{Scale} {Bound}-{Constrained} {Minimization}
  {Problems},'' {\em SIAM J. Sci. Comput.}~{\bf 21}, pp.~1--23, Jan. 1999.

\bibitem{jones_scipy:_2001}
E.~Jones, T.~Oliphant, and P.~Peterson, ``{SciPy}: {Open} source scientific
  tools for {Python},'' {\em http://www. scipy. org/} , 2001.

\bibitem{herraez_fast_2002}
M.~A. Herráez, D.~R. Burton, M.~J. Lalor, and M.~A. Gdeisat, ``Fast
  two-dimensional phase-unwrapping algorithm based on sorting by reliability
  following a noncontinuous path,'' {\em Applied Optics}~{\bf 41}, p.~7437,
  Dec. 2002.

\bibitem{edeson_dimensional_2009}
R.~Edeson, N.~Morris, A.~Tatnall, and G.~S. Aglietti, ``Dimensional {Stability}
  {Testing} on a {Space} {Optical} {Bench} {Structure},'' {\em AIAA
  Journal}~{\bf 47}, pp.~219--228, Jan. 2009.

\bibitem{cook_planetary_2015}
T.~Cook, K.~Cahoy, S.~Chakrabarti, E.~Douglas, S.~C. Finn, M.~Kuchner,
  N.~Lewis, A.~Marinan, J.~Martel, D.~Mawet, B.~Mazin, S.~R. Meeker,
  C.~Mendillo, G.~Serabyn, D.~Stuchlik, and M.~Swain, ``Planetary {Imaging}
  {Concept} {Testbed} {Using} a {Recoverable} {Experiment}–{Coronagraph}
  ({PICTURE} {C}),'' {\em J. Astron. Telesc. Instrum. Syst}~{\bf 1}(4),
  pp.~044001--044001, 2015.

\bibitem{cahoy_wavefront_2013}
K.~L. Cahoy, A.~D. Marinan, B.~Novak, C.~Kerr, T.~Nguyen, M.~Webber,
  G.~Falkenburg, and A.~Barg, ``Wavefront control in space with {MEMS}
  deformable mirrors for exoplanet direct imaging,'' {\em J. Micro/Nanolith.
  MEMS MOEMS}~{\bf 13}(1), pp.~011105--011105, 2013.

\bibitem{spergel_wide-field_2015}
D.~Spergel, N.~Gehrels, C.~Baltay, D.~Bennett, J.~Breckinridge, M.~Donahue,
  A.~Dressler, B.~S. Gaudi, T.~Greene, O.~Guyon, C.~Hirata, J.~Kalirai, N.~J.
  Kasdin, B.~Macintosh, W.~Moos, S.~Perlmutter, M.~Postman, B.~Rauscher,
  J.~Rhodes, Y.~Wang, D.~Weinberg, D.~Benford, M.~Hudson, W.-S. Jeong,
  Y.~Mellier, W.~Traub, T.~Yamada, P.~Capak, J.~Colbert, D.~Masters, M.~Penny,
  D.~Savransky, D.~Stern, N.~Zimmerman, R.~Barry, L.~Bartusek, K.~Carpenter,
  E.~Cheng, D.~Content, F.~Dekens, R.~Demers, K.~Grady, C.~Jackson, G.~Kuan,
  J.~Kruk, M.~Melton, B.~Nemati, B.~Parvin, I.~Poberezhskiy, C.~Peddie,
  J.~Ruffa, J.~K. Wallace, A.~Whipple, E.~Wollack, and F.~Zhao, ``Wide-{Field}
  {InfrarRed} {Survey} {Telescope}-{Astrophysics} {Focused} {Telescope}
  {Assets} {WFIRST}-{AFTA} 2015 {Report},'' {\em ArXiv e-prints}~{\bf 1503},
  p.~3757, Mar. 2015.

\bibitem{brown_green_2012}
R.~A. Brown, ``Green {Day}? {An} {Old} {Mill} {City} {Leads} a {New}
  {Revolution} in {Massachusetts},'' {\em New England Journal of Higher
  Education} , 2012.

\bibitem{the_astropy_collaboration_astropy:_2013}
{The Astropy Collaboration}, T.~P. Robitaille, E.~J. Tollerud, P.~Greenfield,
  M.~Droettboom, E.~Bray, T.~Aldcroft, M.~Davis, A.~Ginsburg, A.~M.
  Price-Whelan, W.~E. Kerzendorf, A.~Conley, N.~Crighton, K.~Barbary, D.~Muna,
  H.~Ferguson, F.~Grollier, M.~M. Parikh, P.~H. Nair, H.~M. Günther, C.~Deil,
  J.~Woillez, S.~Conseil, R.~Kramer, J.~E.~H. Turner, L.~Singer, R.~Fox, B.~A.
  Weaver, V.~Zabalza, Z.~I. Edwards, K.~Azalee~Bostroem, D.~J. Burke, A.~R.
  Casey, S.~M. Crawford, N.~Dencheva, J.~Ely, T.~Jenness, K.~Labrie, P.~L. Lim,
  F.~Pierfederici, A.~Pontzen, A.~Ptak, B.~Refsdal, M.~Servillat, and
  O.~Streicher, ``Astropy: {A} community {Python} package for astronomy,'' {\em
  Astronomy \& Astrophysics}~{\bf 558}, p.~A33, Oct. 2013.

\bibitem{hunter_matplotlib:_2007}
J.~D. Hunter, ``Matplotlib: {A} 2d graphics environment,'' {\em Computing In
  Science \& Engineering}~{\bf 9}(3), pp.~90--95, 2007.

\bibitem{perez_ipython:_2007}
F.~Pérez and B.~Granger, ``{IPython}: {A} {System} for {Interactive}
  {Scientific} {Computing},'' {\em Computing in Science Engineering}~{\bf 9},
  pp.~21--29, May 2007.

\end{thebibliography}
